\documentclass{article}

\usepackage{PRIMEarxiv}

\usepackage[square,sort]{natbib}

\usepackage[utf8]{inputenc} 
\usepackage[T1]{fontenc}    
\usepackage{hyperref}       
\usepackage{url}            
\usepackage{booktabs}       
\usepackage{amsfonts}       
\usepackage{nicefrac}       
\usepackage{microtype}      
\usepackage{lipsum}
\usepackage{fancyhdr}       
\usepackage{graphicx}       
\graphicspath{{media/}}     
\usepackage{underscore}
\usepackage{url,lineno,microtype}
\usepackage[onehalfspacing]{setspace}
\usepackage{subfig}
\usepackage{braket}
\usepackage{amsmath, amssymb}
\usepackage{mathrsfs}
\usepackage{etoolbox}
\usepackage{adjustbox}
\usepackage{cancel}
\usepackage{placeins}
\usepackage{float}
\usepackage{caption}
\usepackage{breakcites}
\usepackage{xfrac}
\usepackage{graphicx}
\usepackage{acro}
\usepackage{graphics}
\usepackage{bm}
\usepackage{ragged2e}
\usepackage{array}
\usepackage{acro}
\usepackage{dutchcal}
\usepackage{tcolorbox}
\usepackage{verbatim}
\usepackage{longtable}
\usepackage{afterpage}
\usepackage{mathtools}
\usepackage{hyperref}

\newtheorem{equationn}{Equation}
\newtheorem{definition}{Definition}
\newtheorem{theorem}{Theorem}

\newtheorem{corollary}{Corollary}
\newtheorem{proposition}{Proposition}
\newtheorem{lemma}{Lemma}

\newcommand\scalemath[2]{\scalebox{#1}{\mbox{\ensuremath{\displaystyle #2}}}}

\makeatletter
\renewcommand*\env@matrix[1][*\c@MaxMatrixCols c]{%
  \hskip -\arraycolsep
  \let\@ifnextchar\new@ifnextchar
  \array{#1}}
\makeatother

\linenumbers

\title{An Extension Of Combinatorial Contextuality For Cognitive Protocols}

\author{
Abdul Karim Obeid, Axel Bruns, Daniel Angus\\
ARC Centre Of Excellence for Automated\\Decision-Making \& Society (ADM+S)\\
Brisbane, QLD, Australia\\
\texttt{\{Abdul Karim Obeid\}obei@qut.edu.au}
\And
Peter Bruza, Catarina Moreira\\
School of Information Systems\\
Queensland University of Technology\\
Brisbane, QLD, Australia
}

\begin{document}
\nolinenumbers
\maketitle
\begin{abstract}
This article extends the combinatorial approach to support the determination of contextuality amidst causal influences. Contextuality is an active field of study in Quantum Cognition, in systems relating to mental phenomena, such as concepts in human memory \citep{aerts2013concepts}. In the cognitive field of study, a contemporary challenge facing the determination of whether a phenomenon is contextual has been the identification and management of disturbances \citep{dzhafarov2016there}. Whether or not said disturbances are identified through the modelling approach, constitute causal influences, or are disregardableas as noise is important, as contextuality cannot be adequately determined in the presence of causal influences \citep{gleason1957measures}. To address this challenge, we first provide a formalisation of necessary elements of the combinatorial approach within the language of canonical causal models. Through this formalisation, we extend the combinatorial approach to support a measurement and treatment of disturbance, and offer techniques to separately distinguish noise and causal influences. Thereafter, we develop a protocol through which these elements may be represented within a cognitive experiment. As human cognition seems rife with causal influences, cognitive modellers may apply the extended combinatorial approach to practically determine the contextuality of cognitive phenomena.
\end{abstract}


\section{Introduction}
\label{sec:Introduction}

Under the assumption that the properties of a system have well established, pre-existing values prior to measurement, contextuality is when the result of a property's measurement is not independent of the co-properties that are measured along with it \citep{peres1991two}. However, a key characteristic of contextuality is its inability to be explained by any causal relationship. Consequently, any experiment that declares the presence of contextuality must remove all doubt that the phenomenon is the result of some causal influence. In Quantum Information Science, this requirement is solely fulfilled by the `No-Disturbance' (ND) condition \citep{ramanathan2012generalized}, which can be experimentally verified by measurement of consistent marginal probabilities, correspondent to the necessary properties. The ND condition was first described in the work of \citet{gleason1957measures}, who determined the basis of the condition from the physical nature of quantum states. Based on Gleason's work, \citet{kochen1975problem} were able to prove a hypothetical system of orthonormal bases in which no deterministic model of outcomes could be realised that was non-contextual.

It cannot be as easily claimed that the ND condition, let alone 'quantum-like' contextuality is inherent to cognitive processes studied within Quantum Cognition. This is because unlike in Quantum Physics, the emergent theories in which the field is grounded do not have any immediate basis in physical properties (i.e., the necessary measurements are mapped to the cognitive experiment by interpretation). For this reason, cognitive modellers have resorted directly to the examination of the associated probabilistic model, for which numerous works have suggested possible frameworks \citep{aerts2013concepts, asano2014violation, bruza2015probabilistic}. What has been consistently demonstrated is that disturbance is unavoidably inherent in the probabilistic outcomes of cognitive experiments, and this causes the failure of the ND condition \citep{dzhafarov2016there}. This is not to say that quantum-like contextuality does not appear within cognitive experiments, as only causal influences falsify the determination of contextuality, and disturbances are not always reduced to causal influences \citep{atmanspacher2019contextuality}. But rather, cognitive modellers require a method to adequately distinguish disturbances that are due to causal influences from those that are due to noise, which constitutes the first challenge of this article.

The second challenge addressed in this article is that the vast majority of literature published on the determination of quantum-like contextuality does not consider the convex decomposition of an associated probabilistic model into a set of deterministic models. In turn, this prevents the identification of causal influences within deterministic models that cancel each other out when aggregated into the combined probabilistic model, as articulated in \citet{yearsley2019contextuality}'s criticism of \citet{cervantes2018snow}.

This article develops an experimental protocol that addresses both of the forementioned challenges by the combinatorial approach of \citet{acin2015combinatorial}. This is realised by two theoretical elements: the Foulis Randall (F-R) product, and the Weighted Fractional Packing Number (WFPN). 

Specifically, the F-R product is relevant to the first challenge, as it expresses all causal constraints of the ND condition. Here it is combined with a method developed by \citet{chaves2015unifying} for assessing the exact amount of causal influence observed in any causal relationship between two observables. In doing so, we realise a process for determining all the disturbances that are only due to causal influences (and not noise) within arbitrary experimental settings, which in turn addresses the first challenge of the article.

With regard to the WFPN, this directly concerns how (non)contextuality is determined in terms of the combinatorial approach, as the cliques enumerated on its graph structure correspond to the deterministic models of the necessary probabilistic model. \citet{acin2015combinatorial} relate constraints by \citet{shannon1956zero} to said cliques, ensuring that their definition of contextuality remains faithful to the convex decomposition of the relative experiment's probabilistic model, and this overcomes the previously mentioned criticisms of \citet{yearsley2019contextuality}. However so, this definition does not yet anticipate the determination of contextuality under the pretense of disturbances, for which it is here adapted, and this addresses the second challenge.

In addressing both of the forementioned challenges, this article produces a novel experimental protocol within the combinatorial approach, for the adequate determination of quantum-like contextuality given the presence of disturbances.

The article proceeds with the following structure: in Section \ref{sec:relatedWork}, we discuss related work; in Section \ref{sec:EPRFramework}, we relate the EPR framework, a seminal example for modelling contextuality that will assist in the understanding of proceeding sections. In Section \ref{sec:CombinatorialApproach}, we detail relevant definitions of the combinatorial approach; in particular, this section motivates the graph structure of the Weighted Fractional Packing Number (WFPN) as the solution to the previously mentioned challenge concerning convex decompositions of probabilistic models. In Section \ref{sec:CausalModelling}, we introduce key aspects of causal models and diagrams, and substantiate that the ND condition is highly restrictive, as well as that \citet{chaves2015unifying}'s method is necessary for its relaxation in determining quantum-like contextuality. In Section \ref{sec:ModellingCombinatorial}, we declare the necessary mappings to integrate the previously mentioned causal modelling techniques with the combinatorial approach. In Section \ref{sec:determiningContextuality}, we prove the main result of the article: a theorem necessary to determine contextuality in the presence of experimental disturbances.

\section{Related Work}
\label{sec:relatedWork}

This work continues upon previous research undertaken by \citet{obeid2021modelling} in the manner of the combinatorial approach, for determining quantum-like contextuality amidst causal influences. To the best of our knowledge, the only other line of research that has addressed the forementioned issues for determining of quantum-like contextuality is that of \citet{jones2019relating}. We perceive that our work is similar in that we consider causal modelling techniques to remedy the issues concerning disturbance, and assume the `no-hidden-influence' principle, as has been articulated in \citet{causalModelApproachContextualityDisturbingSystems}. However, we distinguish our work in that our approach is based within the combinatorial approach of \citet{acin2015combinatorial}, while \citet{jones2019relating}'s approach is based on the probabilistic causal models which are shown to be equivalent to the `Contextuality-by-Default' framework \citet{dzhafarov2016contextuality}.

\section{The `Einstein-Podolsky-Rosen' Framework}
\label{sec:EPRFramework}

In Quantum Cognition, the Einstein-Podolsky-Rosen (EPR) framework \citep{clauser1969proposed} is one that is largely applied among cognitive modellers in investigations of quantum-like contextuality; for this reason, the framework will be used to convey the results of this paper. The framework involves a system of two parties ($A$ and $B$), in which each have an input measurement that may be configured to one of two settings (${}^{\mathrm{ipt}}\!A\,=\,+1$ and ${}^{\mathrm{ipt}}\!A\,=\,-1$ for party $A$; ${}^{\mathrm{ipt}}\!B\,=\,+1$ and ${}^{\mathrm{ipt}}\!B\,=\,-1$ for party $B$). In either setting, an outcome is observed as ${}^{\mathrm{opt}}\!A\,=\,+1$ or ${}^{\mathrm{opt}}\!A\,=\,-1$ for party $A$, or ${}^{\mathrm{opt}}\!B\,=\,+1$ or ${}^{\mathrm{opt}}\!B\,=\,-1$ for party $B$,. Thereafter, a series of experimental trials are sampled to produce four pair-wise distributions (defined by the input measurements) that communicate the probabilistic outcomes of the experiment:

\begin{definition}
Four pair-wise joint distributions generated by the EPR experimental framework:
\begin{align*}
    P(\,{}^{\mathrm{ipt}}\!A=&+1,\,{}^{\mathrm{ipt}}\!B=+1\,),\\
P(\,{}^{\mathrm{ipt}}\!A=&+1,\,{}^{\mathrm{ipt}}\!B=-1\,),\\
P(\,{}^{\mathrm{ipt}}\!A=&-1,\,{}^{\mathrm{ipt}}\!B=+1\,),\\
P(\,{}^{\mathrm{ipt}}\!A=&-1,\,{}^{\mathrm{ipt}}\!B=-1\,)
\end{align*}
\end{definition}

In the adaption of \citet{bruza2015probabilistic}, the systems corresponded to concepts in a bi-amgiguous conceptual combination, such as ``APPLE CHIP''. Each concept would have two senses e.g., ``APPLE'' has a `fruit' or `computer' sense. Measurements corresponded to priming words given to human subjects who then had to interpret the sense of the associated concept: $+1$ would indicate that the interpretation aligns with the prime. For instance, if the prime was `banana', the human subject interpreted `APPLE' in the `fruit' sense. Conversely, $-1$ would denote the dis-alignment between prime and interpretation.

\begin{table}[H]
    \setstretch{1.75}
    \centering
    \begin{tabular}{ c c c }
        & & CHIP \\
        & & $\begin{matrix}[cc]{}^{\mathrm{ipt}}\!B=+1& {}^{\mathrm{ipt}}\!B=-1\\
        \mathrm{(\,potato\,)}&\mathrm{(\,circuit\,)}
        \end{matrix}
        $ \\
        & & $\begin{matrix}[c c c c] +1\;\, & -1\;\, &\;\, +1 &\;\, -1\end{matrix}$ \nonumber \\
        \rotatebox{90}{APPLE} &
            $\begin{matrix}[lc] {}^{\mathrm{ipt}}\!A=+1 & +1 \\\mathrm{(\,banana\,)}& -1 \\  {}^{\mathrm{ipt}}\!A=-1 & +1 \\\mathrm{(\,computer\,)} & -1 \end{matrix}$ & 
            $\begin{pmatrix}[cc|cc] 0.94 & 0.06 & 0.00 & 0.75 \\0.00 & 0.00 & 0.25 & 0.00 \\\hline 0.00 & 0.35 & 0.47 & 0.00 \\ 0.65 & 0.00 & 0.00 & 0.53 \end{pmatrix}$
    \end{tabular}
	\caption[]{\centering Pair-Wise Joint Distributions Of Conceptual Combination ``APPLE CHIP'' \citep{bruza2015probabilistic}}
    \label{tab:conceptualCombinationsBruza}
\end{table}

Contextuality could be determined by a set of inequalities known as the `Bell-CHSH inequalities', that had been previously conceived by \citet{bell1964einstein}. The inequalities summated the statistical correlations of the pair-wise joint distributions.

\begin{equationn}
The Bell-CHSH inequalities define a violation of the linear system of constraints on the correlations of a probabilistic model:
\begin{align*}
0 \,-\, \mathrm{corr}_{\,\scalemath{0.75}{\begin{matrix}{}^{\mathrm{ipt}}\!A\,=\,+1\\[-0.25em]{}^{\mathrm{ipt}}\!B\,=\,+1\end{matrix}}} \,+\, \mathrm{corr}_{\,\scalemath{0.75}{\begin{matrix}{}^{\mathrm{ipt}}\!A\,=\,-1\\[-0.25em]{}^{\mathrm{ipt}}\!B\,=\,+1\end{matrix}}} \,+\, \mathrm{corr}_{\,\scalemath{0.75}{\begin{matrix}{}^{\mathrm{ipt}}\!A\,=\,+1\\[-0.25em]{}^{\mathrm{ipt}}\!B\,=\,-1\end{matrix}}} \,+\, \mathrm{corr}_{\,\scalemath{0.75}{\begin{matrix}{}^{\mathrm{ipt}}\!A\,=\,-1\\[-0.25em]{}^{\mathrm{ipt}}\!B\,=\,-1\end{matrix}}} \,\leq\, 2\\
0 \,+\, \mathrm{corr}_{\,\scalemath{0.75}{\begin{matrix}{}^{\mathrm{ipt}}\!A\,=\,+1\\[-0.25em]{}^{\mathrm{ipt}}\!B\,=\,+1\end{matrix}}} \,-\, \mathrm{corr}_{\,\scalemath{0.75}{\begin{matrix}{}^{\mathrm{ipt}}\!A\,=\,-1\\[-0.25em]{}^{\mathrm{ipt}}\!B\,=\,+1\end{matrix}}} \,+\, \mathrm{corr}_{\,\scalemath{0.75}{\begin{matrix}{}^{\mathrm{ipt}}\!A\,=\,+1\\[-0.25em]{}^{\mathrm{ipt}}\!B\,=\,-1\end{matrix}}} \,+\, \mathrm{corr}_{\,\scalemath{0.75}{\begin{matrix}{}^{\mathrm{ipt}}\!A\,=\,-1\\[-0.25em]{}^{\mathrm{ipt}}\!B\,=\,-1\end{matrix}}} \,\leq\, 2\\
0 \,+\, \mathrm{corr}_{\,\scalemath{0.75}{\begin{matrix}{}^{\mathrm{ipt}}\!A\,=\,+1\\[-0.25em]{}^{\mathrm{ipt}}\!B\,=\,+1\end{matrix}}} \,+\, \mathrm{corr}_{\,\scalemath{0.75}{\begin{matrix}{}^{\mathrm{ipt}}\!A\,=\,-1\\[-0.25em]{}^{\mathrm{ipt}}\!B\,=\,+1\end{matrix}}} \,-\, \mathrm{corr}_{\,\scalemath{0.75}{\begin{matrix}{}^{\mathrm{ipt}}\!A\,=\,+1\\[-0.25em]{}^{\mathrm{ipt}}\!B\,=\,-1\end{matrix}}} \,+\, \mathrm{corr}_{\,\scalemath{0.75}{\begin{matrix}{}^{\mathrm{ipt}}\!A\,=\,-1\\[-0.25em]{}^{\mathrm{ipt}}\!B\,=\,-1\end{matrix}}} \,\leq\, 2\\
0 \,+\, \mathrm{corr}_{\,\scalemath{0.75}{\begin{matrix}{}^{\mathrm{ipt}}\!A\,=\,+1\\[-0.25em]{}^{\mathrm{ipt}}\!B\,=\,+1\end{matrix}}} \,+\, \mathrm{corr}_{\,\scalemath{0.75}{\begin{matrix}{}^{\mathrm{ipt}}\!A\,=\,-1\\[-0.25em]{}^{\mathrm{ipt}}\!B\,=\,+1\end{matrix}}} \,+\, \mathrm{corr}_{\,\scalemath{0.75}{\begin{matrix}{}^{\mathrm{ipt}}\!A\,=\,+1\\[-0.25em]{}^{\mathrm{ipt}}\!B\,=\,-1\end{matrix}}} \,-\, \mathrm{corr}_{\,\scalemath{0.75}{\begin{matrix}{}^{\mathrm{ipt}}\!A\,=\,-1\\[-0.25em]{}^{\mathrm{ipt}}\!B\,=\,-1\end{matrix}}} \,\leq\, 2
\end{align*}
\textit{Note:While there are many expressions of the Bell-CHSH inequalities, the version provided here most closely resembles that taken from \citet{fine1982hidden}.}
\label{def:bellCHSHequaltiies}
\end{equationn}

In the literature, it is common for the maximal statistical correlation of the Bell inequalities to be referred to as the ``Bell parameter'', and is typically denoted as $\mathcal{B}$. For the EPR framework, the Bell parameter would simply be largest L.H.S. value of any of the lines of Equation \ref{def:bellCHSHequaltiies}.

\begin{definition}
For the EPR framework, the Bell parameter $\mathcal{B}$ is defined as the maximal statistical correlation recorded for all pair-wise joint distributions of its input measurements.
\begin{align*}
\mathcal{B} \;=\; \underset{
 \begin{matrix}
 a \,\in\,\{ \,+1, \,-1 \,\}\\
 b \,\in\,\{ \,+1, \,-1 \,\}
 \end{matrix}}{\mathrm{max}} 
 \raisebox{-0.8em}{\scalemath{2}{|}}&\;
 \mathrm{corr}_{\scalemath{0.75}{\begin{matrix}{}^{\mathrm{ipt}}\!A\,=\,+1\\[-0.25em]{}^{\mathrm{ipt}}\!B\,=\,+1\end{matrix}}} \;+\; 
 \mathrm{corr}_{\scalemath{0.75}{\begin{matrix}{}^{\mathrm{ipt}}\!A\,=\,+1\\[-0.25em]{}^{\mathrm{ipt}}\!B\,=\,-1\end{matrix}}}
 \\[-1.5em]
&\;\;\;\;\;\;\;+\;
\mathrm{corr}_{\scalemath{0.75}{\begin{matrix}{}^{\mathrm{ipt}}\!A\,=\,-1\\[-0.25em]{}^{\mathrm{ipt}}\!B\,=\,+1\end{matrix}}} \;+\; \mathrm{corr}_{\scalemath{0.75}{\begin{matrix}{}^{\mathrm{ipt}}\!A\,=\,-1\\[-0.25em]{}^{\mathrm{ipt}}\!B\,=\,-1\end{matrix}}}
\;-\; 2\mathrm{corr}_{\scalemath{0.75}{\begin{matrix}{}^{\mathrm{ipt}}\!A\,=\,a\\[-0.25em]{}^{\mathrm{ipt}}\!B\,=\,b\end{matrix}}}\; \raisebox{-0.8em}{\scalemath{2}{|}}
\end{align*}
\end{definition}

Furthermore, the R.H.S. of Equation \ref{def:bellCHSHequaltiies} (which is 2 for the EPR framework) is known as the classical bound on the statistical correlations of the Bell inequalities. It is the largest value that can be obtained without violating noncontextual hidden variable theories.

\begin{definition}
The value $\mathcal{B}_0$ is defined as the maximal Bell parameter that can be obtained without violating noncontextual hidden variable theories.
\end{definition}

Then, Equation \ref{def:bellCHSHequaltiies} can be simplified to the following expression.

\begin{equationn}
A simplification of Equation \ref{def:bellCHSHequaltiies} that integrates the usage of the Bell parameter.
\begin{align*}
    \mathcal{B} \leq \mathcal{B}_0
\end{align*}
\end{equationn}

For the previously mentioned experiment of \citet{bruza2015probabilistic}, a violation of any of the Bell inequalities would constitute evidence that the concepts are `quantum-like' contextual. While true in principle, this depth of analysis did not identify if the violation of the inequalities was the effect of some causal influence, as pointed out by \citet{dzhafarov2016there}; later in this article, a set of techniques will be introduced in the manner of the combinatorial approach to remediate this challenge. Nevertheless, the elements described in the EPR framework demonstrate an experimental specification necessary to investigate quantum-like contextuality, and will consequently be recalled when introducing various preliminaries used to communicate the findings.

\section{The Combinatorial Approach}
\label{sec:CombinatorialApproach}
The combinatorial approach of \citet{acin2015combinatorial} introduces contextuality scenarios as a hypergraph-based abstraction of a given experiment. The Weighted Fractional Packing Number (WFPN) constitutes the method for determining contextuality within the approach, and it is demonstrated here that the WFPN must be extended in order to support experimental results that exchange disturbances. 

\subsection{Contextuality scenarios}

For any experiment involving the determination of contextuality, contextuality scenarios offer a hypergraph-based abstraction on which all further procedures are conducted.

\begin{definition}
For an experiment of $n$ parties, $H$ denotes a system of contextuality scenarios that correspond to the experiment. The relative contextuality scenario of any party $i$ is then a hypergraph $H_i$, in which its measurements are hyperedges $E(H_i)$, and the possible outcomes of said measurements are vertices $V(H_i)$.
\begin{align*}
H = \{H_1, \ldots, H_n\}, \;\;\; \mathrm{such\;that}\;\;\; E(H_i) \subset 2^{V(H_i)} \;\;\; \mathrm{and} \!\!\!\!\! \bigcup_{\;\;\;\;\;e\, \in\, E(H_i)}\!\!\!\!\!\!\!\!e \;= \;V(H_i)
\end{align*}
\end{definition}

Furthermore, for a hyperedge $e \in E(H_A)$, a vertex $v \in e$ describes an outcome for the measurement, and is notated as $v|e$. For any party, all outcomes and measurements are conventionally reduced to numerical shorthands, which is reflected within their verices and edges - the same also applies for joint outcomes among arbitrary parties. Some shorthands relative to the EPR framework are communicated in Table \ref{tab:shorthands}.

\begin{table}[H]
    \setstretch{1.75}
    \centering
    \begin{center}
    \begin{tabular}{|| l | c ||} 
     \hline
     Outcomes \& Measurements & Shorthand \\ [0.5ex] 
     \hline\hline
     ${}^{\mathrm{opt}}\!A = +1$ & $0\,|$ \\ 
     \hline
     ${}^{\mathrm{opt}}\!A = -1$ & $1\,|$ \\
     \hline
     ${}^{\mathrm{ipt}}\!A = +1$ & $|\,0$\\
     \hline
     ${}^{\mathrm{ipt}}\!A = -1$ & $|\,1$ \\
     \hline
     ${}^{\mathrm{opt}}\!A = +1 \;|\; {}^{\mathrm{ipt}}\!A = +1$ & $0\,|\,0$ \\
     \hline
     ${}^{\mathrm{opt}}\!A = -1, \,{}^{\mathrm{opt}}\!B = +1 \;|\; {}^{\mathrm{ipt}}\!A = +1, \, {}^{\mathrm{ipt}}\!B = +1$ & $1,0\,|\,0,0$ \\
     \hline
     ${}^{\mathrm{opt}}\!A = -1, \,{}^{\mathrm{opt}}\!B = +1 \;|\; {}^{\mathrm{ipt}}\!A = -1, \, {}^{\mathrm{ipt}}\!B = +1$ & $1,0\,|\,1,0$ \\ [1ex] 
     \hline
    \end{tabular}
    \end{center}
	\caption[]{\centering Conventional Shorthands Applied Within The Combinatorial Approach of \citet{acin2015combinatorial}}
    \label{tab:shorthands}
\end{table}

Then, the contextuality scenarios of the parties $A$ and $B$ are hypergraphs $H_A$ and $H_B$.

\begin{definition}
The contextuality scenarios $H_A$ and $H_B$ for the parties $A$ and $B$.
\begin{align*}
    H_A = \{\, (\, v_{0|0}, \; v_{1|0}\,), \; (\, v_{0|1}, \; v_{1|1}\,) \, \} \;\;\;\;\;\; H_B = \{\, (\, v_{0|0}, \; v_{1|0}\,), \; (\, v_{0|1}, \; v_{1|1}\,) \, \}
\end{align*}
\end{definition}

\subsection{Probabilistic models}
\label{sec:probabilisticModels}

Probabilistic models define the probabilities of experimental outcomes taking place, and correspond directly to the vertices of the relative contextuality scenarios.

\begin{definition}
For any party $i$, all possible outcomes that can be generated have weightings attributed to a probabilistic model $p$ that coincides with the relative contextuality scenario $H_i$: Formally, the probabilistic model of any hypergraph $H_i$ is an assignment of a probability to each vertex $v \in V(H_i)$.
\begin{align*}
p : V(H_i) \rightarrow [0,1] \;\;\; \mathrm{such\;that}\;\;\; \forall_{e\,\in\, E(H_i)}\sum_{v\,\in\, e}p(v) = 1
\end{align*}
\end{definition}

\subsection{Compositional products}

Compositional products are operations executed on the hypergraphs of one or more contextuality scenarios, in order to assist the determination of contextuality. Further to this detail, the outcome of a compositional product is also a contextuality scenario.

\subsubsection{The Cartesian product}

The first compositional product detailed here is the Cartesian product.

\begin{definition}
For any system of $n$ contextuality scenarios $H$, the Cartesian product $\bigtimes^{n}_{i=1}H_i$ is defined as having the following vertices and hyperedges.
\begin{align*}
    &\times^{n}_{i=1}V(H_i) \;=\; V(H_1) \times \ldots \times V(H_n)\\
    &\times^{n}_{i=1}E(H_i) \;=\; E(H_1) \times \ldots \times E(H_n)
\end{align*}
\label{lab:verticesHi}
\end{definition}

\begin{figure}[H]
\begin{center}
     \includegraphics[width=0.5\textwidth]{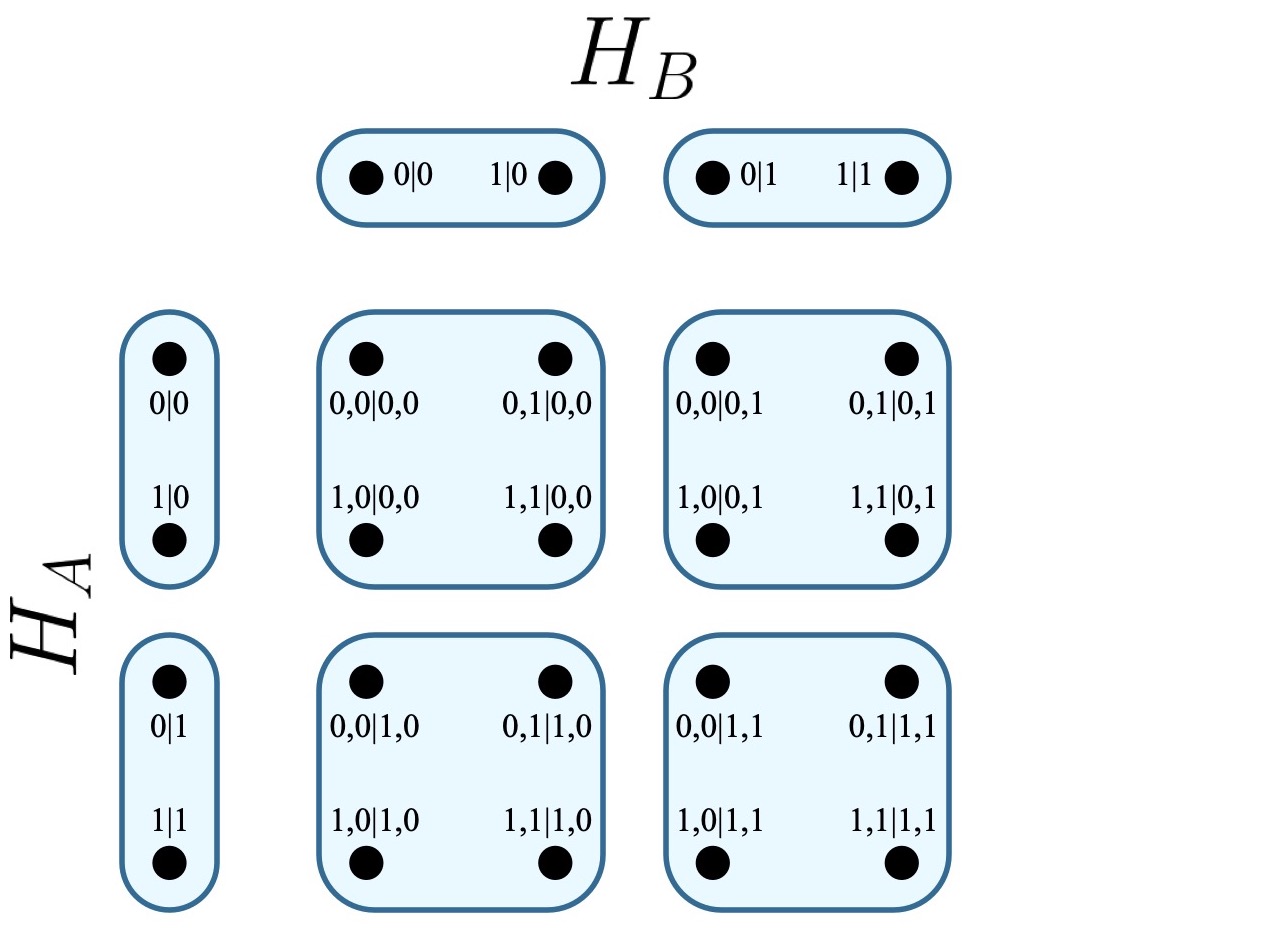}
\end{center}
	\caption{Cartesian Product of Contextuality Scenarios $H_A$ \& $H_B$}
\label{fig:CartesianProduct}
\end{figure}

It follows that any edge of the Cartesian product represents one of the possible combinations of joint measurements that can be conducted between the respective parties of the system. In the EPR experiment, the Cartesian product $H_A \times H_B$ is visualised in Figure \ref{fig:CartesianProduct}.

\subsubsection{Measurement protocols}

In certain cases, it may be necessary to describe the measurements of a given party $j$ as the result of the outcomes of another party $i$. For this purpose, the combinatorial approach defines measurement protocols.

\begin{definition}
A measurement protocol $E_{H_i \rightarrow H_j}$ is a hypergraph generated from a function that maps one or more measurements (as hyperedges) of a contextuality scenario $H_i$ to all the measurements (as hyperedges) of another contextuality scenario $H_j$.
\begin{align*}
    E_{H_i \,\rightarrow\, H_j} := \left\{ \;\bigcup_{v\,\in \,e} \{v\} \times f(v) \;:\; e \in E(H_i), \;\;f : e \rightarrow E(H_j) \;\right\}
\end{align*}
\end{definition}

\begin{figure}[H]
\begin{center}
     \includegraphics[width=0.65\textwidth]{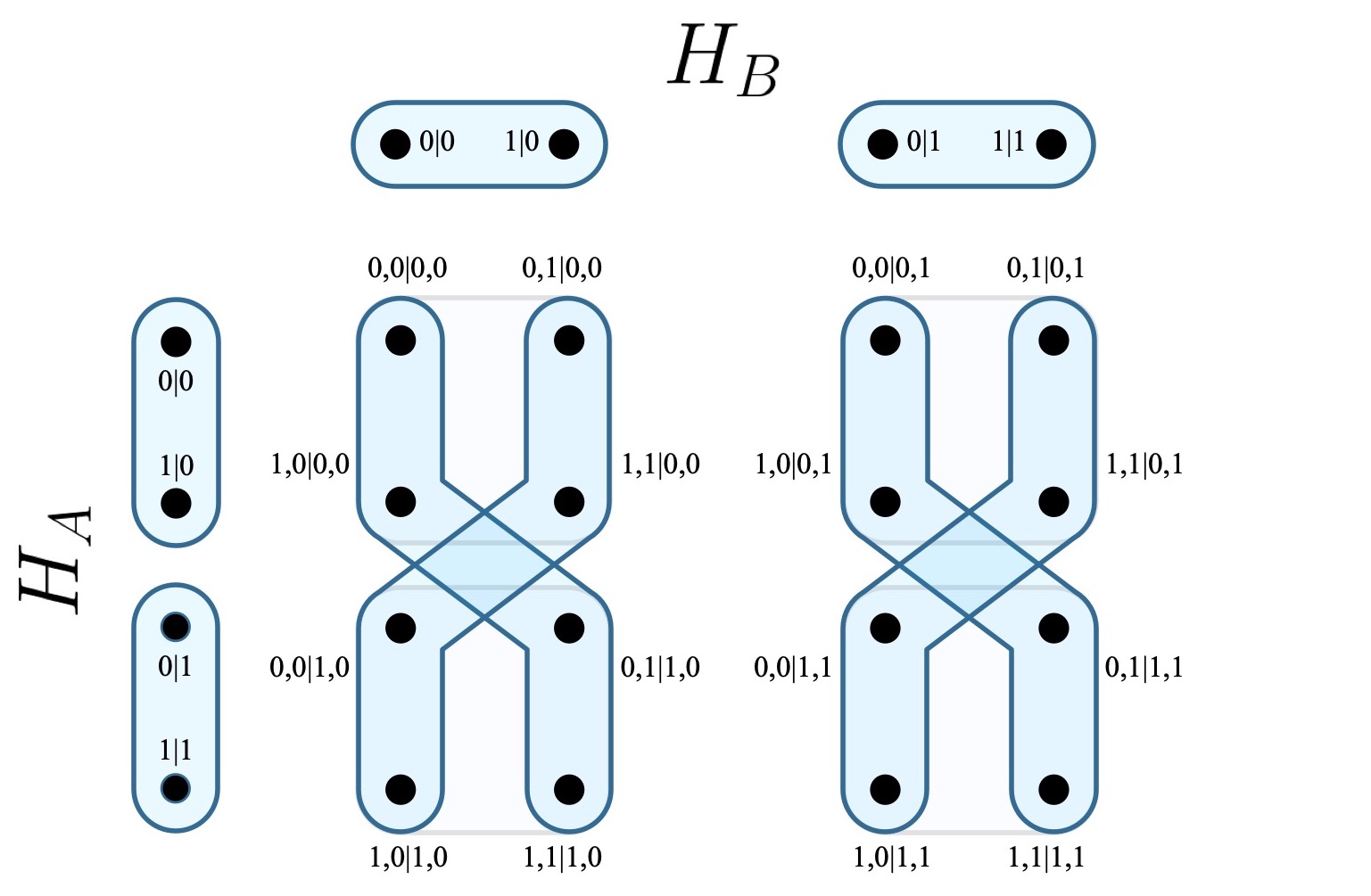}
\end{center}
	\caption{Measurement Protocol $E_{H_A \rightarrow H_B}$}
\label{fig:MeasurementProtocol}
\end{figure}

In terms of an experiment, this is taken to mean that for any outcomes associated with a measurement in $E(H_i)$, that a measurement from $E(H_j)$ is consequently chosen. Recalling the EPR experiment, it is possible that the party $B$ may choose their measurements as a function of $A$'s outcomes. The resulting measurement protocol $E_{H_A \rightarrow H_B}$ visualises this relation in Figure \ref{fig:MeasurementProtocol}.

\subsubsection{The Foulis-Randall product}

By combining the hyperedges that correspond to all measurement protocols of all parties mapped to all other parties, one is able to describe the measurement of any party as the result of any other party's outcome. These are conveyed by hyperedges of the relative contextuality scenarios, whose vertices correspond to the specific outcomes of such events taking place. This comprises the Foulis-Randall (F-R) product, which is yet another compositional product of the combinatorial approach. For a system of $n$ contextuality scenarios $H$, the F-R product is expressed as both a hypergraph and a contextuality scenario, and has numerous variations, all of which containing the same vertices as the Cartesian product (see Definition \ref{lab:verticesHi}). This article is concerned with the \emph{common} F-R product, which is denoted as ${}^{\mathrm{comm}}\bigotimes^{n}_{i=1}H_i$.

\begin{definition}
The hyperedges $E(\,{}^{\mathrm{comm}}\bigotimes^{n}_{i=1}H_i\,)$ of the common F-R product are defined as the union of all possible orderings of the commutative, non-associative `$\otimes$' operator on all contextuality scenarios of a system $H$.
\begin{align*}
    &E(\,{}^{\mathrm{comm}}\bigotimes^{n}_{i=1}H_i\,) \; := \; \left\{ \;e : e \in \left\{ \;\bigotimes^{n}_{i=1}H_i\; \right\}\; \right\}\\&\;\;\;\;\;\;\;\;\;\;\;\;\;\;\;\;\;\;\;\;\;\;\;\;\;\; \mathrm{where} \;\; H_i \otimes H_j  \; := \; \left\{\; E_{H_i \,\rightarrow\, H_j} \;\cup\; E_{H_j \,\rightarrow \,H_i}\; \right\}
\end{align*}
\end{definition}



\begin{figure}[H]
\begin{center}
     \includegraphics[width=0.575\textwidth]{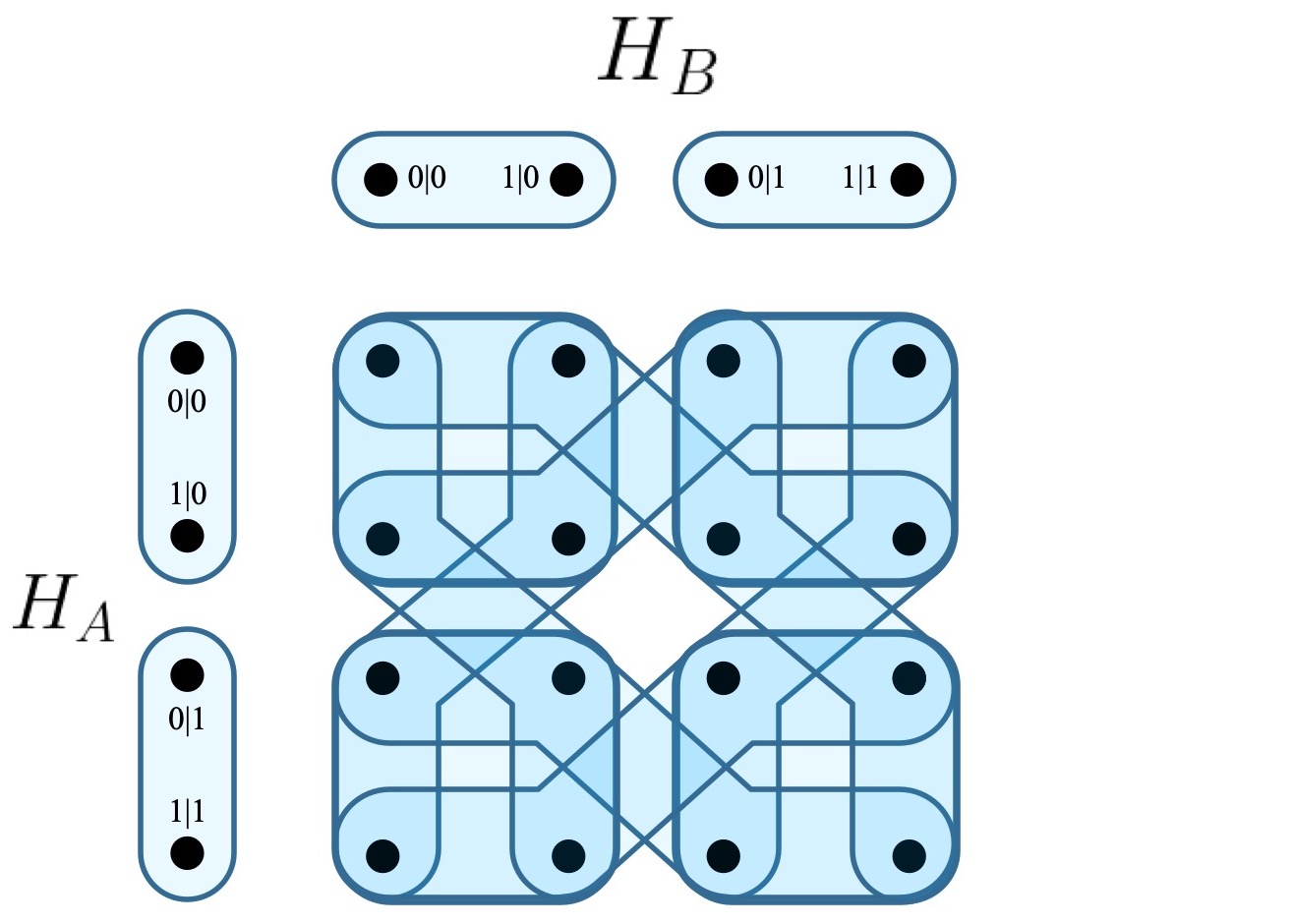}
\end{center}
	\caption{F-R Product Of Contextuality Scenarios $H_A$ \& $H_B$}
\label{fig:FRProduct}
\end{figure}

For the EPR framework, the common F-R product would be equivalent to $H_A \otimes H_B$, which is visualised in Figure \ref{fig:FRProduct}.

\subsubsection{The No Disturbance Condition}

As a widely tested prerequisite for determining contextuality, the No Disturbance (ND) condition ensures that absolutely no causal influences are mutually exchanged between the parties of the given experiment. This satisfies the requirement that contextuality cannot be due to causal influences. The ND condition is achieved by imposing constraints on the probabilistic weightings of the parties. In terms of the EPR framework, the condition is calculated as follows.

\begin{definition}
For an EPR experiment involving two parties, $A$ and $B$, the ND condition requires that given any input measurement by either party, that the probabilistic outcomes must be invariant with respect the outcomes of all other parties.
\begin{align*}
    \mathrm{Pr}(\;{}^{\mathrm{opt}}\!A \,\;|\;\,{}^{\mathrm{ipt}}\!A\;) \;&=\; \mathrm{Pr}(\;{}^{\mathrm{opt}}\!A \,\;|\;\,{}^{\mathrm{ipt}}\!A, {}^{\mathrm{ipt}}\!B = +1\;)\\
&=\; \mathrm{Pr}(\;{}^{\mathrm{opt}}\!A \,\;|\;\,{}^{\mathrm{ipt}}\!A, {}^{\mathrm{ipt}}\!B=-1\;)\\
\mathrm{Pr}(\;{}^{\mathrm{opt}}\!B \,\;|\;\,{}^{\mathrm{ipt}}\!B\;) \;&=\; \mathrm{Pr}(\;{}^{\mathrm{opt}}\!B \,\;|\;\,{}^{\mathrm{ipt}}\!B, {}^{\mathrm{ipt}}\!A = +1\;)\\
&=\; \mathrm{Pr}(\;{}^{\mathrm{opt}}\!B \,\;|\;\,{}^{\mathrm{ipt}}\!B, {}^{\mathrm{ipt}}\!A=-1\;)
\end{align*}
\label{eq:NDConditionAB}
\end{definition}

Within the combinatorial approach, the vertices of the corresponding contextuality scenarios $H_A$ and $H_B$ convey the outcomes for the respective marginal probabilities of all parties. As such, the probabilities of their joint observation with the edges of all other parties restore the necessary expressions of the ND condition.

\begin{definition}
For any outcome corresponding to a vertex $v$ of given party $H_A$ jointly observed with a measurement corresponding to a hyperedge $e$ of a given party $H_B$, the probability is equivalent to that of the same outcome observed with any other measurement corresponding to a hyperedge $e'$ of $H_B$. The same applies with respect to all $H_B$'s outcomes when jointly observed with measurements of $H_A$.
\begin{align*}
    &\sum_{w \, \in \, e}p(v,w) = \sum_{w \, \in \, e'}p(v,w) \;\;\; \forall v \in V(H_A), \;\;\; e, e' \in E(H_B)\\
    &\sum_{w \, \in \, e}p(v,w) = \sum_{w \, \in \, e'}p(v,w) \;\;\; \forall v \in V(H_B), \;\;\; e, e' \in E(H_A)
\end{align*}
\end{definition}

Furthermore, as the F-R product previously defined each of its hyperedges as corresponding to any party's outcome as a function of another party, it follows that the summated probabilities of said hyperedges would need to be collectively equivalent in order to motivate the assumption that no single causal influence is more probable than another.

\begin{corollary}
For all hyperedges of the F-R product that corresponds to the EPR framework $E(\,H_A \,\otimes \,H_B\,)$, the summation of all probabilities of any single hyperedge are equivalent to those of any other hyperedge.
\begin{align*}
    \forall_{e,\,e' \,\in\, E(\,H_A \, \otimes \, H_B\,)} \;\;\sum_{v \, \in \, e}p(v)\; = \!\!\sum_{\;\;v' \, \in \, e'}\!\!\!p(v')
\end{align*}
\label{cly:frProductBalance}
\end{corollary}

\subsubsection{Non-Orthogonality Graphs \& The Weighted Fractional Packing Number}

The WFPN is central to determining contextuality within the combinatorial approach. In order to calculate the Weighted Fractional Packing Number (WFPN), one must first calculate the Non-Orthogonality (NO) graph.

\begin{definition}
The Non-Orthogonality (NO) graph is defined as a simple graph $\mathrm{NO}(\,H\,)$ of the same vertices as an input contextuality scenario $H$, and has hyperedges for vertices that are not within the input's common hyperedges.
\begin{align*}
V(\,\mathrm{NO}(\,H\,)\,) \; :=&\; V(\,H\,) \\
E(\,\mathrm{NO}(\,H\,)\,) \; :=&\; \left \{ \;e : u \sim v \,\Leftrightarrow\, \not\exists e \in E(\,H\,) \;\; \mathrm{such\;that} \;\; \{ \,u,\,v\, \} \,\subseteq\, e \,\right \}\nonumber
\end{align*}
\end{definition}

Of significance is the NO graph of the F-R product, 
$\mathrm{NO}(\,{}^{\mathrm{comm}}\bigotimes^{n}_{i=1}H_i\,)$, for which all maximal cliques $C$ are enumerated.

\begin{definition}
Let $C$ denote all maximal cliques enumerated upon the the NO graph $\mathrm{NO}(\,{}^{\mathrm{comm}}\bigotimes^{n}_{i=1}H_i\,)$.
\begin{align*}
    c \in C \subseteq \mathrm{NO}(\,{}^{\mathrm{comm}}\bigotimes^{n}_{i=1}H_i\,)
\end{align*}
\end{definition}

\begin{definition}
The enumeration of the maximal cliques $C$ exemplify all noncontextual deterministic models, and are indexed by a set of weightings $q$. Then, for any clique $c$, the corresponding model is attributed the weight $q_c$, which is derived from the total number of times that the model occurs within experimentation, as a percentage of the observation of all possible deterministic models. 
\begin{align*}
    \forall_{q_i \,\in\, q} \;\;\; q_i \,\in\, [\,0,\,1\,]
\end{align*}
\end{definition}

In this respect, the WFPN is calculated by constraint of the summation of the weightings $q$.

\begin{definition}
The WFPN $\alpha^{*}$ of the F-R product $\mathrm{NO}(\,{}^{\mathrm{comm}}\bigotimes^{n}_{i=1}H_i\,)$ is equivalent to the summation of all weightings of the set $q$, as indexed by the cliques $C$.
\begin{align*}
\sum_{c \,\in\, C} q_c = \alpha^{*}(\,\mathrm{NO}(\,{}^{\mathrm{comm}}\bigotimes^{n}_{i=1}H_i\,)\,) = 1
\end{align*}
\label{eq:deterministicOnes}
\end{definition}

Finally, the probabilistic model $p$ of the system is recovered by summation of the deterministic models.

\begin{definition}
For a system of contextuality scenarios $H$, the probability $p(v)$ of observing any outcome associated with a vertex $v \in V(H)$ is equivalent to the sum of the deterministic weightings corresponding to the cliques of $C$ that intersect $v$.
\begin{align*}
    \forall_{v \,\in\, V(H)} \;\; p(v) = \sum_{c \;\in\; C}\;\sum_{c \;\cap\; v} q_c
\end{align*}
\label{def:convexDecompQ}
\end{definition}

While the WFPN determines the (non)contextuality of a system, it does so with the assumption that no disturbances are in the experimental results; for this reason, the WFPN must be extended. To do so, we refer to the clique enumeration. As mentioned, this corresponds to all noncontextual deterministic models of the system.

Previous attempts to classify quantum-like contextuality have not taken into account the deterministic decompositions that form their respective probabilistic models: as an implication, experimental disturbances may have influenced the determination of contextuality, despite being overlooked within probabilistic outcomes. Consider an implementation of the EPR framework in which two experimental trials are firstly conducted. Their deterministic models are conveyed by the pair-wise joint distributions in Table \ref{tab:pairwiseJointIndivExpTrials}.

\setcounter{subfigure}{0}
\begin{table}[H]%
    \setstretch{1.75}
    \centering
    \subfloat[\centering Pair-Wise Joint Distributions Of Individual Experimental Trial 1]{{\resizebox{.45\textwidth}{!}{\begin{tabular}{ c c }
        & $\begin{matrix}[cc]{}^{\mathrm{ipt}}\!B=+1\;\;&\;\; {}^{\mathrm{ipt}}\!B=-1\end{matrix}$ \\
        & $\begin{matrix}[ c c c c ] +1 & -1 \;\;&\;\; +1 & -1\end{matrix}$ \nonumber \\
            $\begin{matrix}[cc] {}^{\mathrm{ipt}}\!A=+1 & +1 \\& -1 \\  {}^{\mathrm{ipt}}\!A=-1 & +1 \\ & -1 \end{matrix}$ & 
            $\begin{pmatrix}[cc|cc] \;1.00 & 0.00 & 0.00 & 1.00\;\; \\0.00 & 0.00 & 0.00 & 0.00 \\\hline 1.00 & 0.00 & 0.00 & 0.00 \\ 0.00 & 0.00 & 0.00 & 1.00 \end{pmatrix}$
    \end{tabular}}}}%
    \qquad
    \subfloat[\centering Pair-Wise Joint Distributions Of Individual Experimental Trial 2]{{\resizebox{0.45\textwidth}{!}{\begin{tabular}{ c c }
        & $\begin{matrix}[cc]{}^{\mathrm{ipt}}\!B=+1\;\;&\;\; {}^{\mathrm{ipt}}\!B=-1\end{matrix}$ \\
        & $\begin{matrix}[ c c c c ] +1 & -1 \;\;&\;\; +1 & -1\end{matrix}$ \nonumber \\
            $\begin{matrix}[cc] {}^{\mathrm{ipt}}\!A=+1 & +1 \\& -1 \\  {}^{\mathrm{ipt}}\!A=-1 & +1 \\ & -1 \end{matrix}$ & 
            $\begin{pmatrix}[cc|cc] \;0.00 & 0.00 & 0.00 & 0.00\;\; \\0.00 & 1.00 & 1.00 & 0.00 \\\hline 0.00 & 0.00 & 1.00 & 0.00 \\ 0.00 & 1.00 & 0.00 & 0.00 \end{pmatrix}$
    \end{tabular}}}}%
    \caption{Pair-Wise Joint Distributions Of Two Individual Experimental Trials}%
    \label{tab:pairwiseJointIndivExpTrials}%
\end{table}

After both experimental trials, the probabilistic model is derived from the normalisation of their summated results, as given in Table \ref{tab:pairwiseJointIndivExpTrialsCombined}.

\begin{table}[H]%
    \setstretch{1.75}
    \centering
    \begin{tabular}{ c c }
        & $\begin{matrix}[cc]{}^{\mathrm{ipt}}\!B=+1\;\;&\;\; {}^{\mathrm{ipt}}\!B=-1\end{matrix}$ \\
        & $\begin{matrix}[c c c c] +1 \;\;&\;\; -1 & +1 & -1\end{matrix}$ \nonumber \\
            $\begin{matrix}[cc] {}^{\mathrm{ipt}}\!A=+1 & +1 \\& -1 \\  {}^{\mathrm{ipt}}\!A=-1 & +1 \\ & -1 \end{matrix}$ & 
            $\begin{pmatrix}[cc|cc] \;0.50 & 0.00 & 0.00 & 0.50\;\; \\0.00 & 0.50 & 0.50 & 0.00 \\\hline 0.50 & 0.00 & 0.50 & 0.00 \\ 0.00 & 0.50 & 0.00 & 0.50 \end{pmatrix}$
    \end{tabular}
    \caption{Pair-Wise Joint Distributions Of Combined Experimental Trials From Table \ref{tab:pairwiseJointIndivExpTrials}}%
    \label{tab:pairwiseJointIndivExpTrialsCombined}%
\end{table}

While evaluating the ND condition (Definition \ref{eq:NDConditionAB}) on the probabilistic model of Table \ref{tab:pairwiseJointIndivExpTrialsCombined} does not reveal any disturbances, the same does not hold for the deterministic models in Table \ref{tab:pairwiseJointIndivExpTrials}. This is not to say that the observed disturbances in Table \ref{tab:pairwiseJointIndivExpTrials} constitute causal influences (as Section \ref{sec:associationNDBad} will detail that the ND condition is incorrectly provisioned for this task), but that a significant aspect of the determination of contextuality within cognitive experiments has been so far overlooked. This claim has already been made by \citet{yearsley2019contextuality} and \citet{atmanspacher2019contextuality} for experimentation conducted by \citet{cervantes2018snow}, however a generalisation of the method to retrieve all deterministic models has not yet been considered for cognitive experiments. In fact, this can only be achieved by anticipating all deterministic models (such as those of Table \ref{tab:pairwiseJointIndivExpTrials}) for any arbitrary experiment. Only then can a sensitive treatment of the possible disturbances within the corresponding experimental trials be realised. As detailed in Definition \ref{eq:deterministicOnes}, all deterministic models of an experiment are retrieved by the enumerated cliques of the WFPN. And this holds for any experiment in which the corresponding contextuality scenarios realise an NO graph, demonstrating where the extension of the combinatorial approach should be realised.

For usage later in later sections of this paper, the following definitions are introduced.

\begin{definition}
Let all cliques enumerated by the F-R product of a system of contextuality scenarios $H$ be denoted as ${}^{\mathrm{NC}}C$, as they correspond to the outcomes of all noncontextual deterministic models of the relative system.
\begin{align*}
    {}^{\mathrm{NC}}C \;\subseteq\; \mathrm{NO}(\,{}^{\mathrm{comm}}\bigotimes^{n}_{i=1}H_i\,)
\end{align*}
\end{definition}

\begin{definition}
Let all possible deterministic models for a system $H$ correspond to the cliques ${}^{\mathrm{ALL}}C$ enumerated on the NO graph of the Cartesian product of the system.
\begin{align*}
    {}^{\mathrm{ALL}}C \;\subseteq\; \mathrm{NO}(\,\times^{n}_{i=1}H_i\,)
    \label{eq:allcliques}
\end{align*}
\textit{Note: For a system of contextuality scenarios, all its possible deterministic models correspond to the cliques enumerated on the NO graph of its Cartesian product.}
\end{definition}

By subtracting the cliques ${}^{\mathrm{NC}}C$ from ${}^{\mathrm{ALL}}C$, we derive the cliques that correspond to the deterministic models that are not noncontextual (i.e., either derived from causal influences or noise). The result will be hereafter denoted as ${}^{\cancel{\mathrm{NC}}}C$.

\begin{definition}
The set of cliques ${}^{\cancel{\mathrm{NC}}}C$ that correspond to the deterministic models that are not noncontextual are defined as the difference of the set of cliques ${}^{\mathrm{ALL}}C$ and ${}^{\mathrm{NC}}C$.
\begin{align*}
    {}^{\cancel{\mathrm{NC}}}C \;:=\; {}^{\mathrm{ALL}}C \,\setminus\, {}^{\mathrm{NC}}C
\end{align*}
\textit{Note: Here it is assumed that all cliques derive from operations on the same system of contextuality scenarios.}
\end{definition}

\section{Causal Modelling Techniques}
\label{sec:CausalModelling}

In this section, we detail causal models and diagrams, and argue the failure of the ND condition to distinguish between causal influences and noise for the adequate determination of contextuality amidst causal influences. Furthermore, we relate the work of \citet{chaves2015unifying} for correctly quantifying causal influences.

\subsection{Causal Models \& Diagrams}

In certain cases, cognitive experiments may appear to determine contextuality by probabilistic weightings, but are in fact noncontextual due to hidden causal influences. To identify said causal influences, one must apply causal modelling techniques, and as such, causal models are here established as necessary. \citet{pearl2009causality} defines any causal model as follows.

\begin{definition}
A causal model $\mathrm{M}$ is an ordered triple, in which $\mathrm{U}$ represent the set of exogenous random variables, $\mathrm{V}$ represent the set of endogenous random variables, and $\mathrm{E}$ is the set of causal influences: formally the expressions of the values of $\mathrm{V}$ as functions of the values within $\mathrm{U}$ and $\mathrm{V}$.
\begin{align*}
    \mathrm{M} \;=\; \langle \,\mathrm{U}, \mathrm{V}, \mathrm{E}\, \rangle
\end{align*}
\end{definition}

For use in further equations, the set of exogenous and endogenous variables will be unified into a single set.

\begin{definition}
The set $\mathrm{X}$ denotes the union of the sets $\mathrm{U}$ and $\mathrm{V}$.
\begin{align*}
    \mathrm{X} := \{ \, \mathrm{U} \, \cup \, \mathrm{V} \, \}
\end{align*}
\end{definition}

Furthermore, causal models described here assume a canonical probabilistic model (distinct from Section \ref{sec:probabilisticModels}) attributes weightings to all $n$ members of the set $\mathrm{X}$.

\begin{definition}
A canonical probabilistic model assigns probabilistic weightings to all random variables $\mathrm{X}$ of a causal model.
\begin{align*}
    \forall_{\mathrm{x} \,\in\, \mathrm{X}} \;\;\; \mathrm{Pr}(\,\mathrm{x}\,) \rightarrow [\,0,\,1\,] \;\;\;\; \mathrm{such\;that} \;\;\; \mathrm{Pr}(\; \mathrm{X}_1 , \, \ldots, \, \mathrm{X}_n \; ) = 1
\end{align*}
\end{definition}

Causal diagrams provide a visual abstraction of causal models, by ascribing the random variables to a simple graph.

\begin{definition}
A causal diagram ascribes the random variables within the set $\mathrm{X}$ to nodes of a directed acyclic graph $\mathcal{G}$. Therein, the directed edges $\mathcal{E}$ correspond to the causal influences established by the set of expressions $\mathrm{E}$.
\begin{align*}
    \mathcal{G} \;=\; (\,\mathrm{X},\, \mathcal{E}\,)
\end{align*}
\end{definition}

\begin{definition}
The function $f_{\mathrm{pnt}}$ defines the causal influences, for which any $f_{\mathrm{pnt}}(\,\mathrm{x}\,)$ of a variable $\mathrm{x}$ returns the parent vertex of $\mathrm{x}$ within the causal diagram. This defines the set of directed edges $\mathcal{E}$.
\begin{align*}
    f_{\mathrm{pnt}}(\,\mathrm{x}\,) \;\subset\; \mathrm{X}\, \cup \{ \, \varnothing \,\} \;\;\;\; \mathrm{such\;that} \;\;\;\; \forall \mathrm{x}' \,\in \, f_{\mathrm{pnt}}(\,\mathrm{x}\,) \;\; \exists(\,\mathrm{x}', \,\mathrm{x}\,) \in \mathcal{E}
\end{align*}
\label{def:fparent}
\end{definition}

\begin{equationn}
The joint distribution factors of the random variables within the set $\mathrm{X}$ are defined as the product of all probabilities, given observation of the variables associated with their parent nodes in the corresponding causal diagram.
\begin{align*}
    \mathrm{Pr}(\,\mathrm{X}\,) \;=\; \prod_{i}\mathrm{Pr}( \; \mathrm{X}_i \; | \; f_{\mathrm{pnt}}(\,\mathrm{X}_i\,) \; )
\end{align*}
\label{eq:distributionFactors}
\end{equationn}

\begin{definition}
Let the set of all exogenous variables be collected into a latent variable $\Lambda$.
\begin{align*}
    \Lambda \; := \; \{ \; \mathrm{x} : \mathrm{x} \in \mathrm{X} \;\; \mathrm{and} \;\; f_{\mathrm{pnt}}(\,\mathrm{x}\,) = \{\,\varnothing\,\}   \; \}
\end{align*}
\end{definition}

\subsection{Determining Causal Influences}

\subsubsection{Association}
\label{sec:associationNDBad}

Failures of purely probabilistic attempts to classify cognitive experiments as determining contextuality are due to modellers relying upon the lowest level, \textit{association}, of \citet{pearl2018book}'s ladder of causation in order to identify causal influences. 

\begin{equationn}
Association proposes a causal influence from a variable $\mathrm{Z}$ to another variable $\mathrm{Y}$ by fulfilment of the following probabilistic expression.
\begin{align*}
    \mathrm{Pr}(\,\mathrm{Y}\, | \,\mathrm{Z}\,) > \mathrm{Pr}(\,\mathrm{Y}\, | \,\neg\mathrm{Z}\,)
\end{align*}
\label{eq:association}
\end{equationn}

This expression can be evaluated by probabilistic weightings alone. At best, association only proves that $\mathrm{Y}$ \emph{could} be caused by $\mathrm{Z}$, vice versa, that both are caused by some other variable, or may be due to noise within experimental results.

It is important to note that while association does not prove any cause for either $\mathrm{Y}$ and $\mathrm{Z}$, that it nevertheless is the primary clue that causation may be at play. It follows that imposing constraints upon associations between variables not only restrict all forms of causation, but the experimental noise that is necessary to determine contextuality. Such constraints happen to form the basis of the ND condition.

\begin{proposition}
The ND condition constitutes a system of expressions that substitute the operator of Equation \ref{eq:association}.
\end{proposition}
\textit{\textbf{Proof:} Replace the observation of $\mathrm{Y}$ with ${}^{\mathrm{opt}}\!A$, the observation of $\mathrm{Z}$ with ${}^{\mathrm{ipt}}\!A, {}^{\mathrm{ipt}}\!B = -1$, and the observation of $\neg\mathrm{Z}$ with ${}^{\mathrm{ipt}}\!A, {}^{\mathrm{ipt}}\!B = +1$ in Equation \ref{eq:association}. Then substitute the operator with that of equivalence (i.e., `=') to recover an expression of Definition \ref{eq:NDConditionAB}. The same is achieved for any other operation within the ND condition for any combination of inputs.}$\hfill\blacksquare$

\subsubsection{Intervention}

In order to determine contextuality for cognitive experiments that may be due to noise, it is necessary to restrict only causal influences. As such, one must refer to the second level, \textit{intervention}, of \citet{pearl2018book}'s ladder of causation. 

\begin{equationn}
Intervention for a causal influence from some variable $\mathrm{Z}$ to another variable $\mathrm{Y}$ requires that the experimenter deliberately fixes the protocols that both exhibit and inhibit $\mathrm{Z}$, as denoted by the $\textit{do}$ operator on a probabilistic expression.
\begin{align*}
    \mathrm{Pr}(\,\mathrm{Y}\, | \,\mathrm{\it{do}}(\mathrm{\,Z\,})\,) > \mathrm{Pr}(\,\mathrm{Y}\, | \,\mathrm{\it{do}}(\mathrm{\,\neg Z\,})\,)
\end{align*}
\label{eq:doExpression}
\end{equationn}

This expression clearly determines whether $\mathrm{Y}$ is caused by $\mathrm{Z}$, and cannot be conceptualised by probabilistic weightings alone. Intervention has been detailed in the framework of \citet{chaves2015unifying}, as the necessary technique to differentiate causal influences from noise. And so two equations from the framework are here recalled:

\begin{equationn}
Given intervention on all $n$ variables $\mathrm{X}$ of a causal model that are observed under pretense of their parent nodes, the $\textit{do}$ operator redefines the joint distribution factors as follows.
\begin{align*}
    \mathrm{Pr}(\,\mathrm{X}\,|\,\textit{do}(\,\mathrm{X}_i = k\,)\,) = \begin{cases}
\prod^n_{j \,\neq\, i}\mathrm{Pr}(\,\mathrm{X}_j \,|\,f_{\mathrm{pnt}}(\,\mathrm{X}_j\,)\,) & \mathrm{if}\;\;\mathrm{X}_i  \neq k \\ 
 0 & \mathrm{otherwise}
\end{cases}
\end{align*}
\label{eq:revisedJointDistributionFactors}
\end{equationn}

The second equation\footnote{Refer to Equation 4 of \citet{chaves2015unifying}.} simply augments Equation \ref{eq:doExpression} with respect to the latent variable $\Lambda$, to determine the degree to which $Z$ causally influences $Y$. In this paper, the expression is generalised.

\begin{definition}
The expression $\mathrm{C}_{{}_{\mathrm{X}_i = k, \,\mathrm{X}_j = k'}}$ concerns two variables, $\mathrm{X}_i$ and $\mathrm{X}_j$ respectively, and returns the direct causal influence that some outcome $\mathrm{X}_i = k$ has upon some other outcome $\mathrm{X}_j = k'$.
\begin{align*}
    \mathrm{C}_{{}_{\mathrm{X}_i = k, \,\mathrm{X}_j = k'}} = \!\!\!\!\!\!\! \underset{\scalemath{0.75}{\begin{matrix}\mathrm{X}_j\, =\, k',\\[0.15em]f_{\mathrm{pnt}}(\mathrm{X}_j),\\[0.15em]\mathrm{X}_i \,=\, k,\; \mathrm{X}_i \, \neq \, k\end{matrix}}}{\mathrm{sup}} \!\! \sum_{\lambda \,\in\, \Lambda}\mathrm{Pr}(\,\lambda\,)|\mathrm{Pr}(\,\mathrm{X}_j = k'\, |\,do(\, \mathrm{X}_i = k\,),\, f_{\mathrm{pnt}}(\mathrm{X}_j), \, \lambda\,)\;\;\;\;&\\[-3em]
    -\; \mathrm{Pr}(\,\mathrm{X}_j = k'\, |\,do(\, \mathrm{X}_i \neq k\,), \, f_{\mathrm{pnt}}(\mathrm{X}_j), \, \lambda\,)|&\\[-0.5em]
\end{align*}
\label{def:causalInfluenceDO}
\end{definition}

\section{Mapping The Combinatorial Approach To Causal Models}
\label{sec:ModellingCombinatorial}

In order for the combinatorial approach to determine contextuality in the presence of experimental disturbances, this section formalises the necessary causal modelling techniques described in the previous section within the combinatorial approach.

\subsection{Contextuality Scenarios}
\label{sec:contextualityScenariosToRVs}

Contextuality scenarios are the fundamental element of the combinatorial approach, and so require a consistent mapping to the random variables of causal models in order to leverage causal modelling techniques. This is no trivial task, as there is no single mapping between them. To demonstrate this point, two separate alternatives will be detailed here in the manner of the EPR framework. The first alternative is motivated from hidden variable theories that precede the discovery of contextuality. 

\begin{figure}[H]%
    \centering
    \subfloat[Causal Diagram]{{\includegraphics[width=0.25\textwidth]{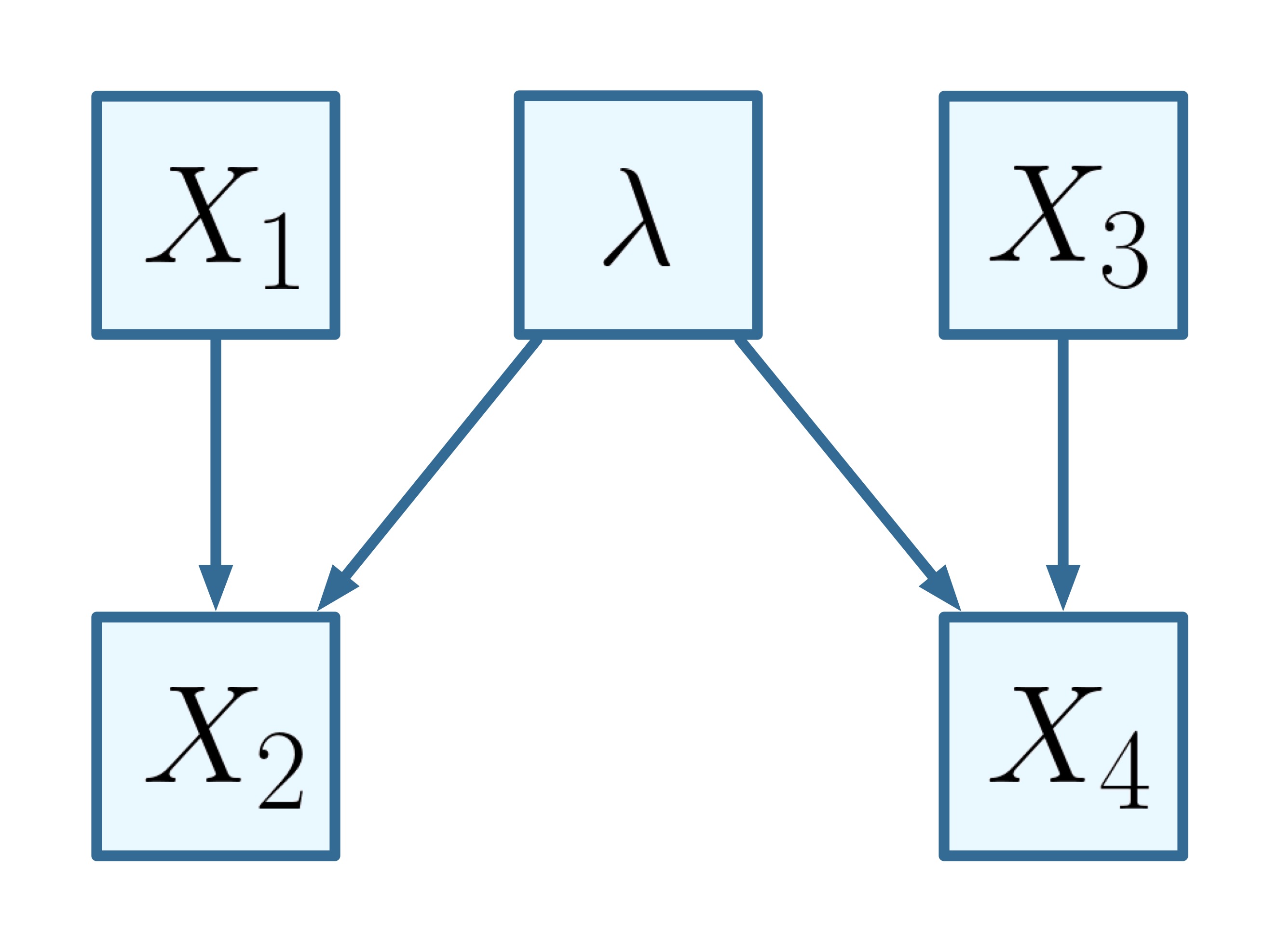}}}%
    \\
    \subfloat[Mapping From Causal Model To Contextuality Scenarios]{{\includegraphics[width=0.75\textwidth]{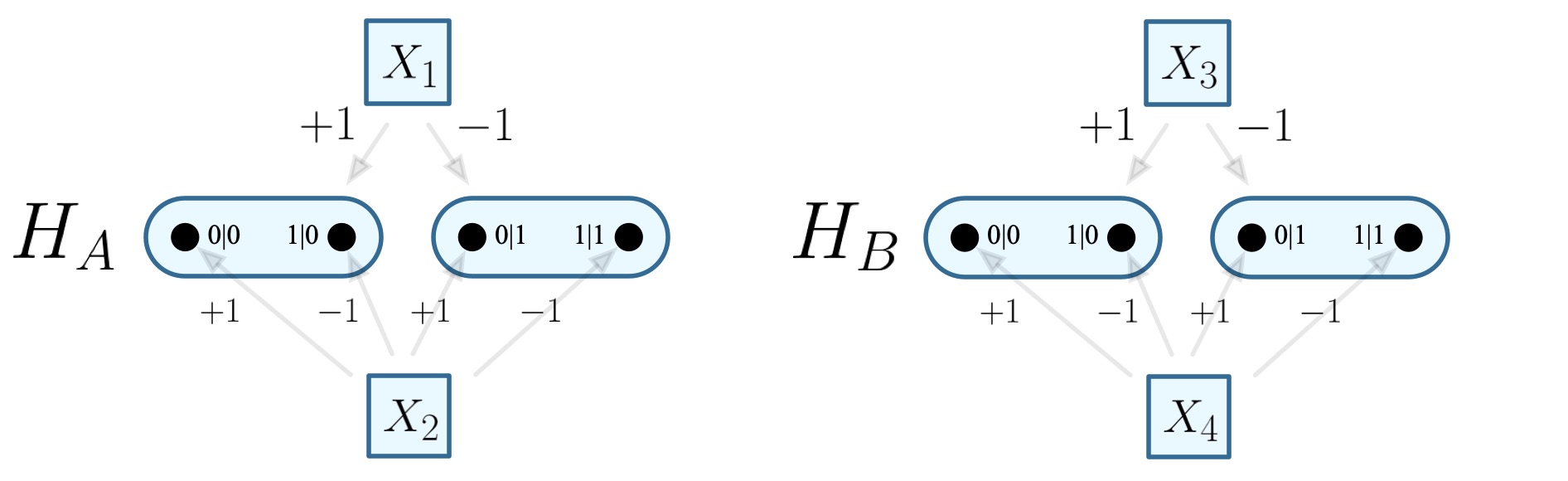}}}%
    \caption{Alternative Mapping Of Causal Model To Contextuality Scenarios With Functional Assignments}%
    \label{fig:CausalModelAlternative1}%
\end{figure}

Recalling the EPR framework's experimental settings, consider that each configuration of both parties' input measurements (${}^{\mathrm{ipt}}\!A\,=\,+1$ and ${}^{\mathrm{ipt}}\!A\,=\,-1$ for party $A$; ${}^{\mathrm{ipt}}\!B\,=\,+1$ and ${}^{\mathrm{ipt}}\!B\,=\,-1$ for party $B$) within the experiment maps to the states of two distinct random variables $\mathrm{X}_1$ and $\mathrm{X}_3$. Thereafter, the outcomes observed (as either ${}^{\mathrm{opt}}\!A\,=\,+1$ or ${}^{\mathrm{opt}}\!A\,=\,-1$ for party $A$, or ${}^{\mathrm{opt}}\!B\,=\,+1$ or ${}^{\mathrm{opt}}\!B\,=\,-1$ for party $B$) for either measurement are also distinct random variables $\mathrm{X}_2$ and $\mathrm{X}_4$. where $f_{\mathrm{pnt}}(\,\mathrm{X}_2\,) =  \{\,\mathrm{X}_1\,\}$ and $f_{\mathrm{pnt}}(\,\mathrm{X}_4\,)  =  \{\,\mathrm{X}_3\,\}$. This alternative assumes that the outcomes observed by either party are functionally assigned by the configuration of the measurements, and has a causal probabilistic model that is visualised in Figure \ref{fig:CausalModelAlternative1}.

\setcounter{subfigure}{0}

The second alternative does not assume functional assignments between random variables, meaning that all causal influences extend from exogenous influences. It implicates a separate interpretation of the experiment: that each dichotomous outcome pertains to its own random variable. In turn, it generates an entirely different system of contextuality scenarios, in which each element of the Cartesian product (which was previously given in Figure \ref{fig:CartesianProduct}) has its own random variable.

While both of the forementioned alternatives are entirely possible, the selection depends upon the researcher's own preferences for how physical measurements and outcomes should map to random variables. As measurements and outcomes constitute the edges and vertices of contextuality scenarios, it follows that this will determine the mapping from random variables to contextuality scenarios.

\begin{definition}
For a system of the contextuality scenarios $H$, the function $f_{\mathrm{vtc}}$ defines a mapping from an outcome $X_i \,=\, k$ to the set of vertices $e'$ that define it within $H$. 
\begin{align*}
    f_{\mathrm{vtc}}(\,\mathrm{X}_i = k\,) \;:\; (\,\mathrm{X}_i = k\,) \rightarrow w \;\;\; \mathrm{where}\;\;\; \,w\,\subseteq V(H)
\end{align*}
\label{eq:mappingFunctionOutcomes}
\end{definition}

\subsection{Causal Influences}
\label{sec:causalInfluences}

As mentioned in Definition \ref{def:fparent}, causal influences are formalised in causal diagrams as directed edges between random variables.

\begin{figure}[H]%
    \centering
    \subfloat[$H_A$ Measurement To $H_A$ Outcome]{{\includegraphics[width=0.25\textwidth]{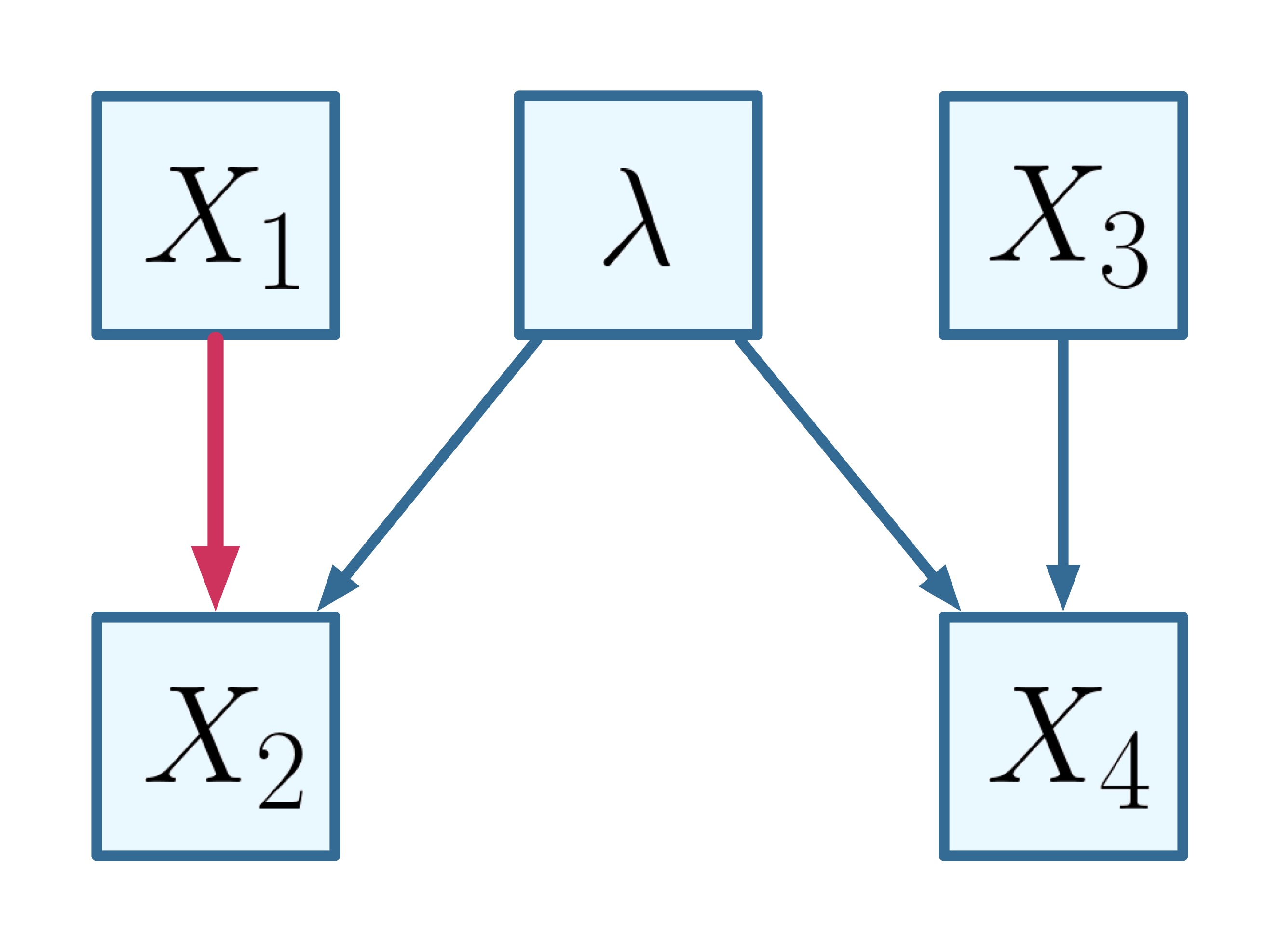}}}%
    \qquad
    \subfloat[$H_A$ Measurement To $H_B$ Outcome]{{\includegraphics[width=0.25\textwidth]{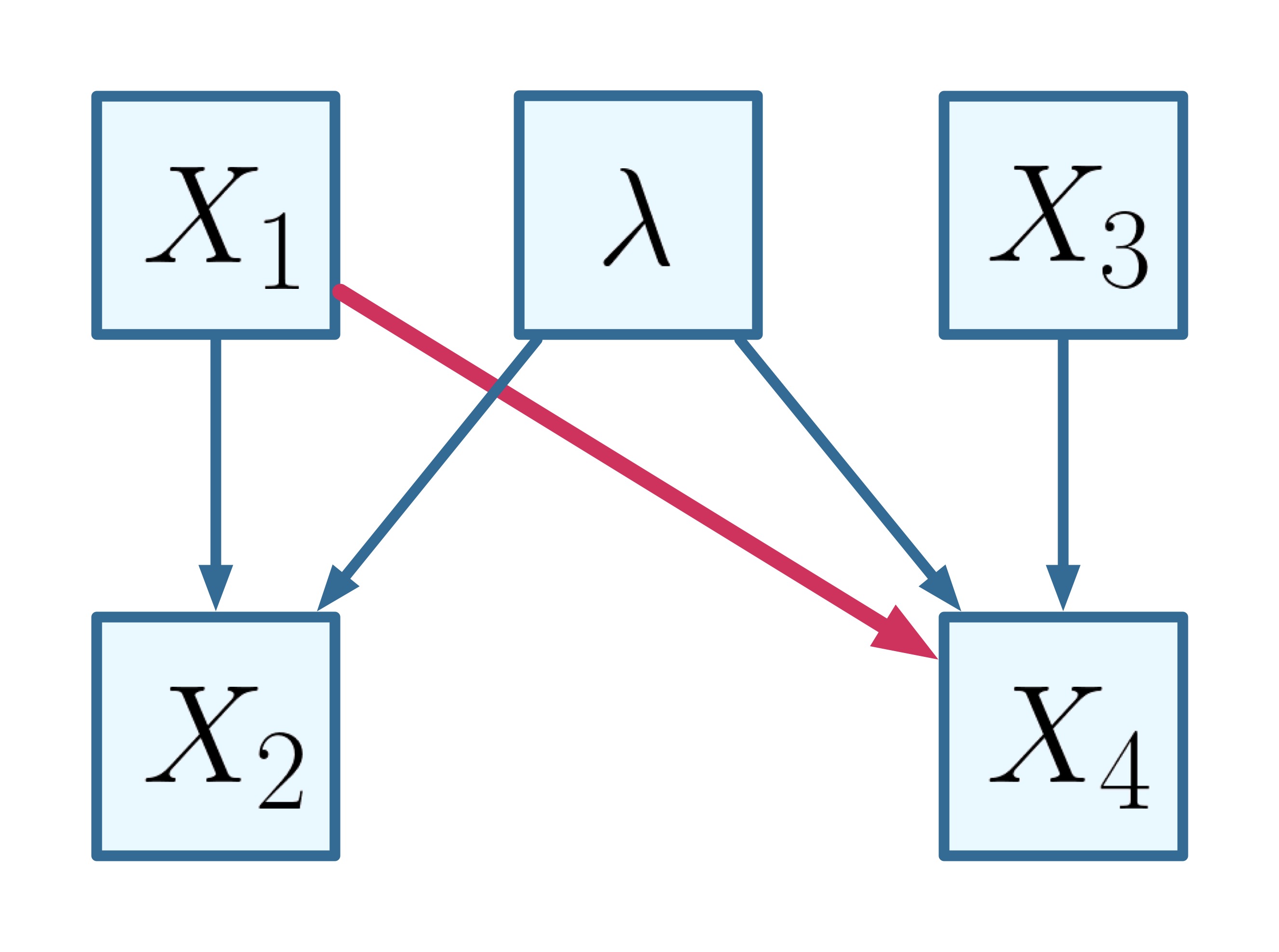}}}%
    \qquad
    \subfloat[$H_A$ Outcome To $H_B$ Outcome]{{\includegraphics[width=0.25\textwidth]{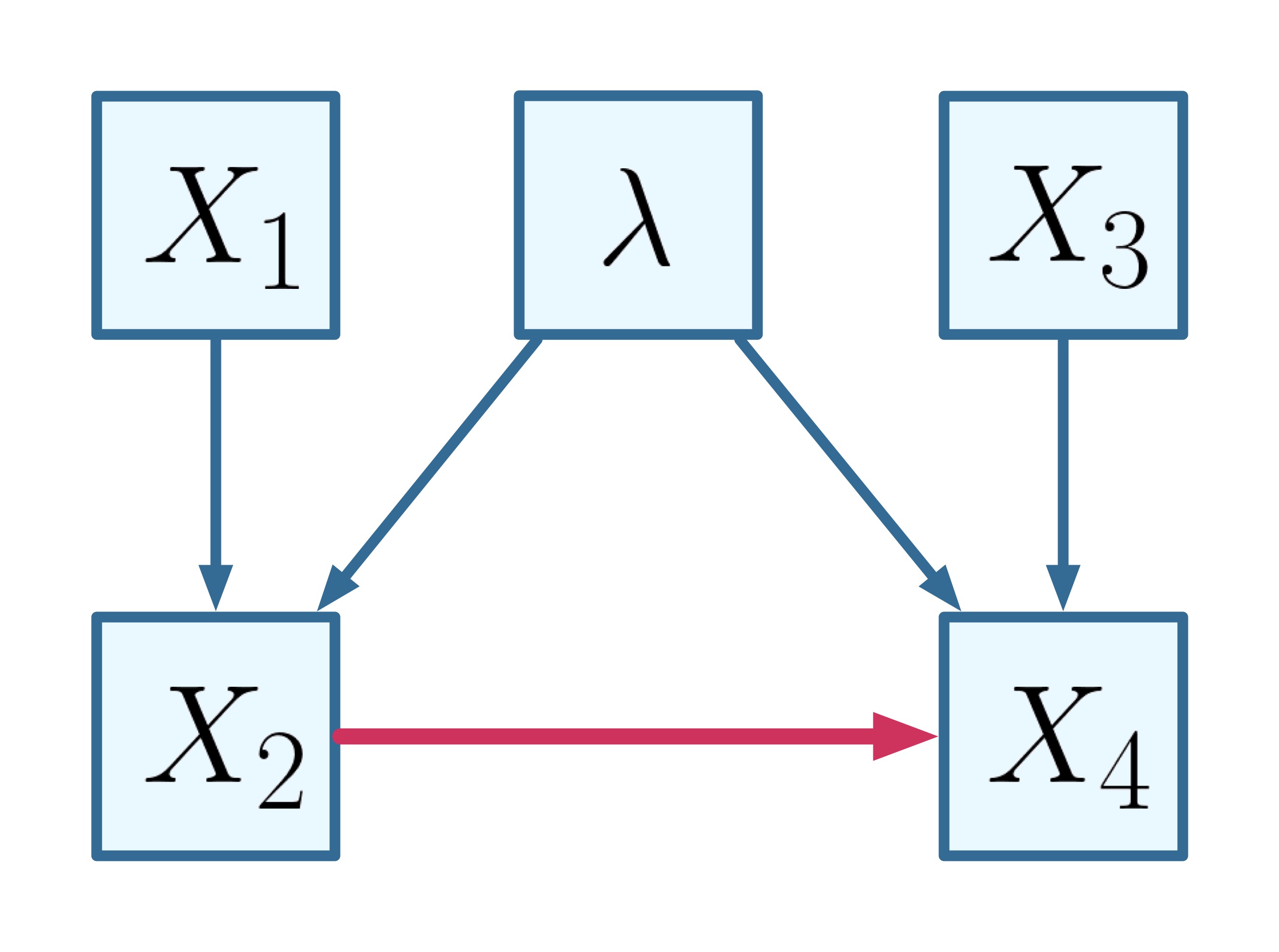}}}%
    \caption{Alternative Causal Influences}%
    \label{fig:CausalModelAlternativeRelations}%
\end{figure}

In terms of contextuality scenarios (and following on from the specification of Section \ref{sec:contextualityScenariosToRVs}), the definition corresponds to a relation between two disjoint subsets of the set of vertices $V(H)$ for a system of contextuality scenarios $H$. This is further specified by the nature of the relation within the causal diagram, for which some possibilities are visualised in the manner of the EPR framework, as shown in Figure \ref{fig:CausalModelAlternativeRelations}.

\setcounter{subfigure}{0}

\begin{figure}[H]
\begin{center}
     \includegraphics[width=0.65\textwidth]{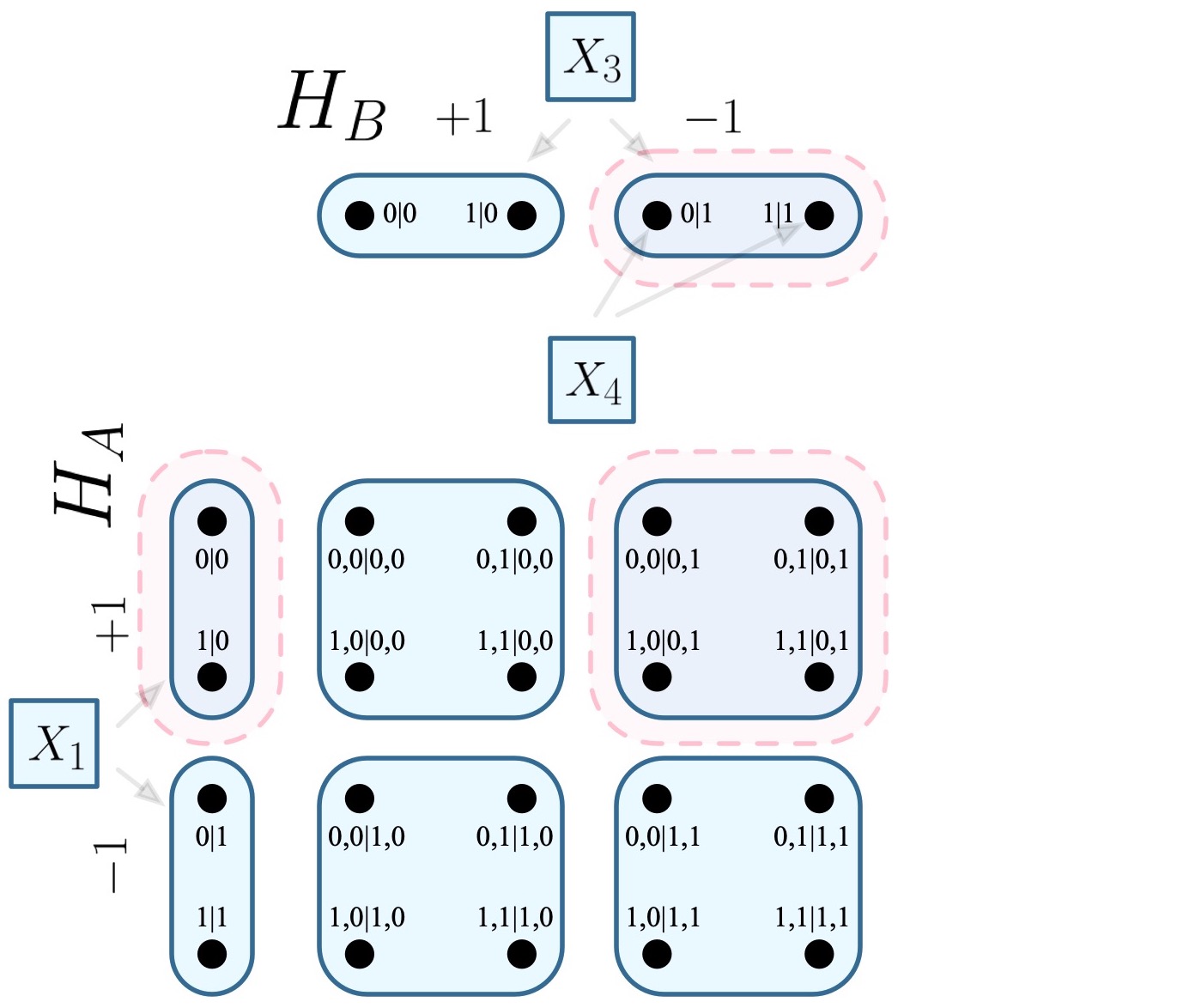}
\end{center}
	\caption{Visualisation Of Relation Within Figure \ref{fig:CausalModelAlternativeRelations}c, Given That $X_3 = -1$}
	\textit{Note: The relation produced by the causal influence (and its derivative hyperedges) are highlighted in red.}
\label{fig:CausalInfluenceAToB}
\end{figure}

Considering that Figure \ref{fig:CausalModelAlternativeRelations}c conveys the choice of a measurement of one party influencing some outcomes on another party, the combinatorial approach defines these as edges and vertices respectively. Following Figure \ref{fig:CausalModelAlternativeRelations}c, suppose one wishes to interrogate whether the highlighted causal influence (i.e., that $X_1$ influences $X_4$), given that $X_3 = -1$. The relation is visualised within the necessary contextuality scenarios in Figure \ref{fig:CausalInfluenceAToB}.

In this respect, the full set of causal influences will hereafter be attributed a set $R$, in which each member details the pair of disjoint vertices that correspond to it.

\begin{definition}
For a causal diagram $\mathcal{G} = \{ \mathrm{X},\,\mathcal{E}\,\}$, and a system of contextuality scenarios $H$, let an edge $e \in \mathcal{E}$ have the source vertex $\mathrm{X}_{i}$, and the destination vertex $\mathrm{X}_{j}$. Then the outcomes $\mathrm{X}_{i} = k$ and $\mathrm{X}_{j} = k'$ defines a member $r \in R$.
\begin{align*}
    \forall_{r \,\in\, R} \;\;\; r \;:=\; \{\,(\,f_{\mathrm{vtc}}(\,\mathrm{X}_i = k\,), \,f_{\mathrm{vtc}}(\,\mathrm{X}_j = k'\,)\,)\,\}
\end{align*}
\end{definition}

\subsection{Probabilities}

As both contextuality scenarios and random variables can share probabilistic assignments, their respective interpretations of probabilistic models are related here.

\begin{equationn}
Any outcome $p(v)$ of the probabilistic model $p$ associated with a system of contextuality scenarios $H$ has an equivalence to the probability of observing one or more of the set of $n$ random variables $\mathrm{X}$.
\begin{align*}
    &\forall_{H_i \,\in\, H} \; \forall_{e\, \in\, E(\,H_i\,)} \; \forall_{v \, \in \, e} \\ &\;\;\;\;\;\;\; p(\,v\,) \;=\; \mathrm{Pr}(\,\mathrm{X}_j,\, \ldots,\, \mathrm{X}_k\,|\,\mathrm{X}_l,\, \ldots,\, \mathrm{X}_m\,) \;\;\; \mathrm{where} \;\;\; j,k,l,m \in [\,1,\,n\,]
\end{align*}
\label{eq:ptoP}
\end{equationn}

By Definition \ref{eq:mappingFunctionOutcomes} and Equation \ref{eq:ptoP}, it is possible to recover the probability of observing any single outcome $\mathrm{X}_i = k$.

\begin{equationn}
The probability of observing any single outcome $\mathrm{X}_i = k$ is equivalent to the summation of the probabilities of all vertices within the set returned by the function $f_{\mathrm{vtc}}(\,\mathrm{X}_i = k\,)$
\begin{align*}
    \!\!\!\!\!\!\!\!\!\mathrm{Pr}(\,\mathrm{X}_i = k\,) \;= \!\!\!\!\!\!\!\!\!\!\!\!\!\!\!\!\!\!\!\!\!\!\!\sum_{\;\;\;\;\;\;\;\;\;\;\;\;\;\;\;\;\;\;\;v\, \in\, f_{\mathrm{vtc}}(\,\mathrm{X}_i \,=\, k\,)}\!\!\!\!\!\!\!\!\!\!\!\!\!\!\!\!\!\!\!\!\!\!\!p(\,v\,)
\end{align*}
\end{equationn}

Similarly, the probability of observing any outcome $\mathrm{X}_i = k$, given the observation of any other outcome $\mathrm{X}_j = k'$ is given as follows.

\begin{equationn}
The probability of observing any outcome $\mathrm{X}_i = k$, given the observation of any other outcome $\mathrm{X}_j = k'$ is equivalent to the summation of the probabilities of all vertices within the set returned by the product of the functions, $f_{\mathrm{vtc}}(\,\mathrm{X}_i = k\,)$ and $f_{\mathrm{vtc}}(\,\mathrm{X}_j = k'\,)$.
\begin{align*}
    \!\!\!\!\!\!\!\!\!\!\!\!\!\!\!\!\!\!\!\!\mathrm{Pr}(\,\mathrm{X}_i = k \;|\; \mathrm{X}_j = k' \,) \;= \!\!\!\!\!\!\!\!\!\!\!\!\!\!\!\!\!\!\!\!\!\!\!\!\!\!\!\!\!\!\!\!\!\!\!\!\!\!\!\!\!\!\!\!\!\!\!\!\!\!\!\!\!\!\!\!\!\!\!\!\!\!\!\sum_{\;\;\;\;\;\;\;\;\;\;\;\;\;\;\;\;\;\;\;\;\;\;\;\;\;\;\;\;\;\;\;\;\;\;\;\;\;\;\;\;\;\;\;\;\;\;\;\;\;\;v\, \in\, \{ \,f_{\mathrm{vtc}}(\,\mathrm{X}_i\, =\, k\,) \,\times\, f_{\mathrm{vtc}}(\,\mathrm{X}_j\, =\, k'\,) \,\}} \!\!\!\!\!\!\!\!\!\!\!\!\!\!\!\!\!\!\!\!\!\!\!\!\!\!\!\!\!\!\!\!\!\!\!\!\!\!\!\!\!\!\!\!\!\!\!\!\!\!\!\!\!\!\!\!\!\!\!\!\!\!\!p(\,v\,)
\end{align*}
\end{equationn}

Furthermore, the joint distribution factors are also formalised, relative to the system of contextuality scenario as follows.

\begin{equationn}
Given a variable $\mathrm{X}_i$ that has been fixed to the outcome $k$, the joint distribution factors are expressed in terms of the relative system of contextuality scenarios $H$. Specifically, the joint distribution only calculates the product over any vertices within $V(H)$ that intersect the corresponding vertices of the outcome $f_{\mathrm{vtc}}(\,\mathrm{X}_i = k\,)$.
\begin{align*}
    \mathrm{Pr}(\,\mathrm{X} \;|\; do(\,\mathrm{X}_i = k\,) \,) \;=\; \begin{cases} \prod_{\,v \,\in\, V(H) }\,p(\,v\,)\;&\mathrm{if} \;\; \{\,v \,\cap f_{\mathrm{vtc}}(\,\mathrm{X}_i = k\,)\,\} \;\not=\;\{\,\varnothing\,\} \\ 0&\mathrm{otherwise} \end{cases}
\end{align*}
\end{equationn}

\newpage

\section{Determining Quantum-Like Contextuality}
\label{sec:determiningContextuality}

In this section, we integrate the elements of the combinatorial approach. and related causal modelling techniques in order to declare the main result of the article: a theorem for determining contextuality within experiments that exchange causal influences.

\subsection{Disturbances In Deterministic Models}

For any deterministic model that corresponds to a clique $c$, any edge $e$ of the F-R product always intersects it at exactly one vertex if any only if the model is noncontextual.

\begin{lemma}
For any deterministic model on a system of contextuality scenarios $H$ whose corresponding clique $c$ is of the set of cliques ${}^{\mathrm{NC}}C$, the cardinality of the set defined by the intersection of the clique and any edge of the F-R product is always equivalent to 1.
\begin{align*}
   \forall_{c \,\in\, {}^{\mathrm{NC}}C} \;\; \forall_{e\, \in \,E(\,\bigotimes^{n}_{i=1}H_i\,)} \;\;\; |\{ \, c \, \cap \, e \, \}| = 1
\end{align*}
\label{lma:noncontextualdeterminism}
\end{lemma}

\textit{\textbf{Proof:} Firstly, it is known that all hyperedges of the Cartesian product of a system of contextuality scenarios correspond to joint distributions of outcomes. Since any deterministic model only records a single deterministic weighting in any joint distribution of outcomes, it follows that any hyperedge within the Cartesian product will only intersect the clique of the correspondent deterministic model once. Secondly, by Corollary \ref{cly:frProductBalance} the summated probabilities corresponding to the vertices of any one hyperedge of the F-R product are exactly equivalent to one another. As the F-R product is a superset of the Cartesian product, it follows that any hyperedge of the F-R product will also intersect any clique once, in order to remain consistent with the subset of hyperedges that form the Cartesian product. Furthermore, to prove that Lemma \ref{lma:noncontextualdeterminism} only holds for noncontextual deterministic models, recall that deterministic models that are not noncontextual must violate the ND condition, which imbalances the equality of Corollary \ref{cly:frProductBalance} by the F-R product. Therefore, it holds that only noncontextual deterministic models can record a single deterministic weighting of 1 in any hyperedge of the F-R product.}$\hfill\blacksquare$

Then, Lemma \ref{lma:noncontextualdeterminism} can be integrated into the calculation of disturbance for all deterministic models that are not noncontextual, by summating the weightings of $q$ that fail the equality.

\begin{lemma}
Consider a system of contextuality scenarios $H$, and the set of weightings $q$ that correspond to all deterministic models indexed by the set of cliques ${}^{\cancel{\mathrm{NC}}}C$. For any single clique $c \, \in \,{}^{\cancel{\mathrm{NC}}}C$, its integration with Lemma \ref{lma:noncontextualdeterminism} defines the deterministic weight that constitutes its disturbance. Furthermore, the summation of all such weightings define the absolute total disturbance for exchanged within the system.
\begin{align*}
    \sum_{\;\;\;c\, \in\, {}^{\cancel{\mathrm{NC}}}C} \!\!\!\!\!\!\!\!\!\!\!\!\!\!\!\!\!\sum_{\;\;\;\;\;\;\;\;\;\;\;\;\;\;\;e\, \in \,E(\,\bigotimes^{n}_{i=1}H_i\,)} \!\!\!\!\!\!\!\!\!\!\!\!\!\!\!\!\!\!q_c\,||\{ \, c \, \cap \, e \, \}| - 1|
\end{align*}
\label{lma:disturbanceCalculated}
\end{lemma}

\begin{figure}[H]
\begin{center}
     \includegraphics[width=0.5\textwidth]{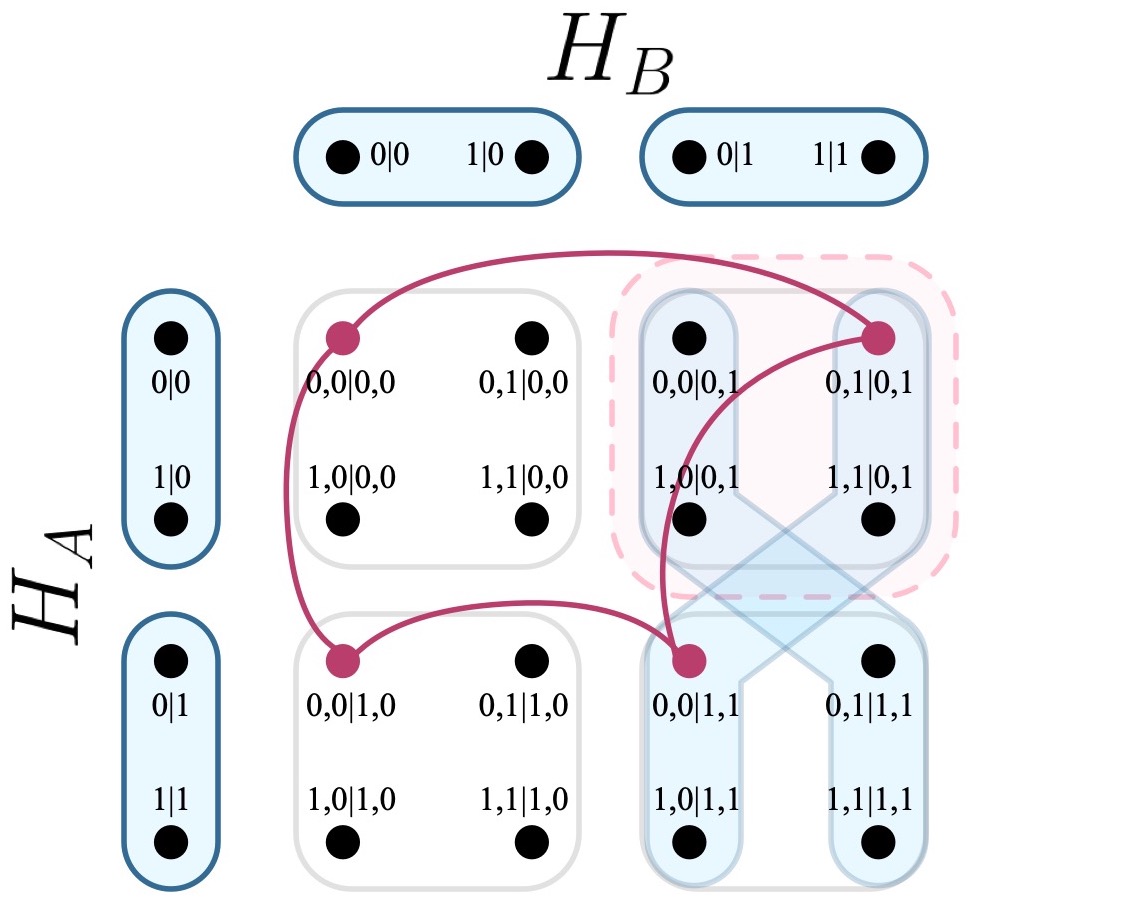}
\end{center}
	\caption{Intersection Of A Relation, Two Edges Of The F-R Product, \& A Clique}
	\textit{Note: As shown above, the clique $c \,\in\, {}^{\cancel{\mathrm{NC}}}C$ (deep red) intersects one edge of the F-R product (light blue) twice, and the other not once. Integrated into the expression $|\{\,c \,\cap \,e \,\}| - 1$, both edges produce the values $+1$ and $-1$ respectively.}
\label{fig:CliqueWeight}
\end{figure}

\textit{\textbf{Proof:} It is known already from Lemma \ref{lma:noncontextualdeterminism} that for a system of contextuality scenarios $H$, that any member $c\, \in \,{}^{\mathrm{NC}}C$ intersects any hyperedge of the F-R product in one vertex. Naturally, it follows that $\forall_{c\, \in \,{}^{\mathrm{NC}}C}$ that $|\{\,c \,\cap \,e \,\}| - 1$ must always equal zero. In fact, irrespective of all other elements of Lemma \ref{lma:disturbanceCalculated}, if ${}^{\cancel{\mathrm{NC}}}C$ were substituted with ${}^{\mathrm{NC}}C$, its clear by the previous point that the entire expression would resolve to zero, validating the fact that deterministic models corresponding to the set ${}^{\mathrm{NC}}C$ do not exchange disturbance. The same should not hold for ${}^{\cancel{\mathrm{NC}}}C$, which is supported by the fact that $|\{\,c \,\cap \,e \,\}| - 1$ does not equal zero for members of ${}^{\cancel{\mathrm{NC}}}C$. In a simple case, such as the system that corresponds to the EPR framework, $|\{\,c \,\cap \,e \,\}| - 1$ can be either $+1$ or $-1$ for any member of ${}^{\cancel{\mathrm{NC}}}C$, which is visualised in Figure \ref{fig:CliqueWeight}.}

\begin{figure}[H]
\begin{center}
     \includegraphics[width=0.65\textwidth]{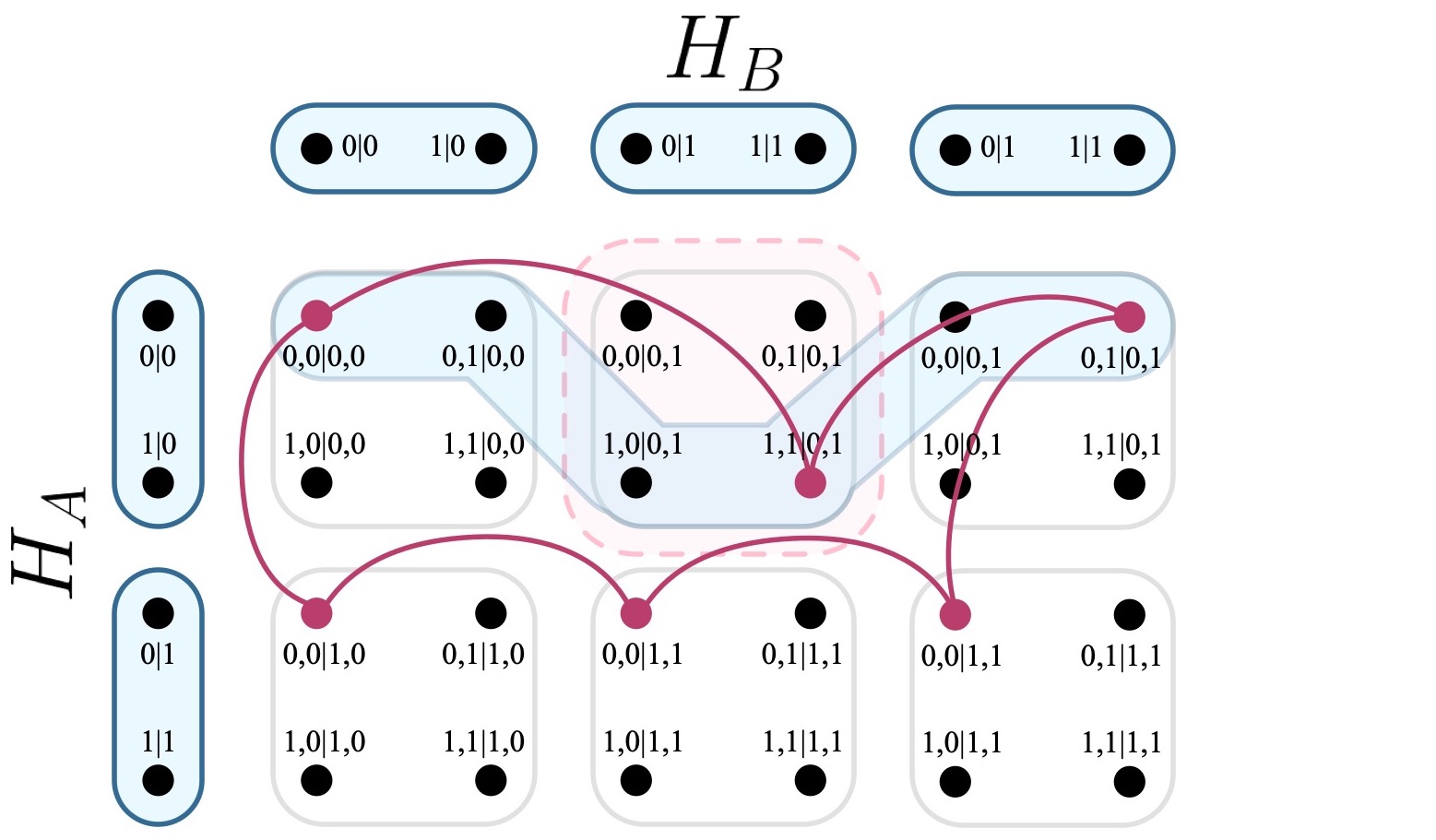}
\end{center}
	\caption{Intersection Of A Relation, An Edges Of The F-R Product, \& A Clique Within A System That Contains Three Measurements}
	\textit{Note: As shown above, the clique intersects the edge of the F-R product thrice, which evaluates the expression $|\{\,c \,\cap \,e \,\}| - 1$ to the value $+2$.}
\label{fig:DisturbanceOnCliqueThreeMeasurements}
\end{figure}

\textit{If the expression $(|\{\,c \,\cap \,e \,\}| - 1) = \pm 1$ were true for arbitrary systems beyond the EPR framework, the expression $\sum_{c \in {}^{\cancel{\mathrm{NC}}}C}1$ would detail the total disturbance. However there exist cases where $(|\{\,c \,\cap \,e \,\}| - 1) > 1$, specifically for deterministic models that exchange disturbance for multiple measurements. Suppose for the EPR framework that the second party were to have three measurements instead of two. A clique $c\, \in\, {}^{\cancel{\mathrm{NC}}}C$ could violate the ND condition for two separate outcomes. This would mean that there are two outcomes that exchange disturbance for this model, as visualised in Figure \ref{fig:DisturbanceOnCliqueThreeMeasurements}}.

\textit{In saying this, the assumption that the disturbance (corresponding to any $c\, \in\, {}^{\cancel{\mathrm{NC}}}C$) is equivalent to 1 would no longer hold. Instead, to correctly quantify disturbances for both outcomes, the ND condition would need to be applied. This is handled by the F-R product, and so Lemma \ref{lma:disturbanceCalculated} accounts for the total disturbance exchanged within a system.}$\hfill\blacksquare$

While Lemma \ref{lma:disturbanceCalculated} defines the expression necessary to calculate the total disturbance exchanged within a system, consider isolating the disturbance for only a single set of measurements from one party onto a set of outcomes of another party. By specification of relations in Section \ref{sec:causalInfluences} and Corollary \ref{cly:frProductBalance}, this is achieved by firstly articulating the edges of the F-R product that capture disturbance for any single relation. 

\begin{definition}
For any relation $r \in R$ that corresponds to any causal influence for a system of contextuality scenarios $H$, the edges of the F-R product necessary for quantifying disturbances on said relation are defined as $E_r$, and are exactly those edges within the necessary measurement protocol that intersect the Cartesian product of the relation.
\begin{align*}
    E_r \;:=\; \left \{ \,e \,:\, e\, \in \,E_{\,H_i \,\rightarrow \,H_j\,} \;\; \mathrm{where} \;\; x \subseteq H_j,\; y \subseteq H_i, \;\; r \,:=\, \{\,(\,x, \,y\,)\,\}, \;\; \mathrm{and}\;\; \left \{\, e \, \cap \, \times^{|r|}_{i=1} r \,\right \} \neq \{\,\varnothing\,\} \, \right \}
\end{align*}
\label{def:ER}
\end{definition}
Then by Definition \ref{def:ER}, one can augment Lemma \ref{lma:disturbanceCalculated} to only quantify weightings that intersect the relation that correspond to the measurements and outcomes in question. Specifically for any probabilistic model, the disturbance on any hyperedge of the F-R product can be evaluated and quantified for whether it corresponds to a relation by the following function.

\begin{lemma}
Consider a probabilistic model $p$, a relation $r \in R$, and an edge $e \in E(\,\otimes^{n}_{i=1}\,)$. Let it be such that all deterministic weightings of the set $q$ form part of the convex decomposition of $p$, as articulated in Definition \ref{def:convexDecompQ}. Then the total disturbance exchanged between the measurements and outcomes of the relation quantified on the said edge within the probabilistic model are returned by the following function.
\begin{align*}
f_{\mathrm{dtb}}(\, r, \, e,\,p\,) \;:= 
\begin{cases} 
\sum_{c\, \in {}^{\cancel{\mathrm{NC}}}C} \;\;q_c\!\left|\left|\left\{ \, c \, \cap \, e\, \right\}\right| - 1\right| & \mathrm{if} \;\;\; e \,\in\, E_r \\
0 &\mathrm{otherwise}
\end{cases}
\end{align*}
\end{lemma}
\textit{\textbf{Proof:} It is known that the Cartesian product of any relation $r$ contains all vertices of its corresponding measurements and outcomes. As any disturbances recorded on said measurements and outcomes can only occur for weightings of $q$ attributed to their respective vertices, it holds that imposing the intersection of $\times^{|r|}_{i=1} r$ by the set $E_r$ on the expression of Lemma \ref{lma:disturbanceCalculated} isolates only the disturbances for said measurements and outcomes. Note that this may include disturbances both to and from the intended parties, as the Cartesian product is agnostic to causal influences. As such, the evaluating hyperedges are fixed to a subset of the F-R product, specifically only those which fit the measurement protocol of the intended source party to the intended destination party. This ensures that only the intended disturbances are captured.}$\hfill\blacksquare$

\subsection{Quantifying Causal Influences}

After calculating the total disturbance on a relation, it is possible to finally formalise Definitiion \ref{def:causalInfluenceDO} within the combinatorial approach.

\begin{definition}
Consider two relations $r'$ and $r''$ which correspond to the observation of the outcome $\mathrm{X}_i\,=\,k$ over another outcome $\mathrm{X}_j \,= \,k'$.
Let it be such that when $\mathrm{X}_i = k$, that the relative probabilistic model is $p'$, and that when $\mathrm{X}_i \neq k$, that the relative probabilistic model is $p''$. Then the direct causal influence from $\mathrm{X}_i$ to $\mathrm{X}_j$ is equivalent to the total disturbance observed when $\mathrm{X}_i = k$ is fixed minus the total disturbance observed when $\mathrm{X}_i \neq k$ is fixed.
\begin{align*}
&\mathrm{C}_{{}_{\mathrm{X}_i\, =\, k, \,\mathrm{X}_j\, =\, k'}} \; = \!\!\!\!\!\!\!\!\!\!\!\!\!\!\!\!\!\!\!\!\!\!\sum_{\;\;\;\;\;\;\;\;\;\;\;\;\;\;\;e\, \in \,E(\,\bigotimes^{n}_{i=1}H_i\,)} \!\!\!\!\!\!\!\!\!\!\!\!\!\!\!\!\!\!\left|\,f_{\mathrm{dtb}}(\, r', \, e,\,p'\,) - f_{\mathrm{dtb}}(\, r'', \, e,\,p''\,)\,\right|\\
&\;\;\;\;\;\;\;\;\;\;\;\;\mathrm{given}\;\;\;r'\;=\; \{\,(\,f_{\mathrm{vtc}}(\,do(\,\mathrm{X}_i\,=\,k\,)\,), \,f_{\mathrm{vtc}}(\,\mathrm{X}_j \,= \,k'\,)\,)\,\}\\
&\;\;\;\;\;\;\;\;\;\;\;\;\;\;\;\;\;\;\;\;\;\;\;\; \mathrm{and}\;\;\;r''\;=\; \{\,(\,f_{\mathrm{vtc}}(\,do(\,\mathrm{X}_i\,\neq\,k\,)\,), \,f_{\mathrm{vtc}}(\,\mathrm{X}_j \,= \,k'\,)\,)\,\}
\end{align*}
Furthermore, as there is a direct correspondence between the measurements and outcomes, $\mathrm{X}_i\, =\, k$ and $\mathrm{X}_j\, =\, k'$, within $\mathrm{C}_{{}_{\mathrm{X}_i\, =\, k, \,\mathrm{X}_j\, =\, k'}}$ and the relation $r'$, it holds that the shorthand $\mathrm{C}_{r'}$ can be used to denote it. Generally, the direct causal influence for any relation $r$ can be denoted by $\mathrm{C}_r$.
\begin{align*}
    \mathrm{C}_{r} \,=\, \mathrm{C}_{{}_{\mathrm{X}_i\, = \,k, \,\mathrm{X}_j\, =\, k'}} \;\;\; \mathrm{where} \;\;\; r \;:=\; \{\,(\,f_{\mathrm{vtc}}(\,\mathrm{X}_i\,=\,k\,), \,f_{\mathrm{vtc}}(\,\mathrm{X}_j \,= \,k'\,)\,)\,\}
\end{align*}
\label{def:CausalVariable}
\end{definition}

Having formalised the direct causal influence for any relation within the combinatorial approach, it is then possible to derive for any experiment the total disturbance that constitutes causal influences.

\begin{lemma}
For any relation $r \, \in \,R$ that corresponds to any causal influence for a system of contextuality scenarios, the total direct causal influences that constitute all or part thereof the total disturbances are equivalent to the following expression.
\begin{align*}
    \sum_{r \,\in\, R}\,\mathrm{D}_r - \mathrm{max}\,\{\,0, \, \mathrm{D}_r - \mathrm{C}_r \,\} \;\;\; \mathrm{where} \;\;\; \mathrm{D}_r = \!\!\!\!\!\!\!\!\!\!\!\!\!\!\!\!\!\!\!\!\!\! \sum_{\;\;\;\;\;\;\;\;\;\;\;\;\;\;\;e\, \in \,E(\,\bigotimes^{n}_{i=1}H_i\,)} \!\!\!\!\!\!\!\!\!\!\!\!\!\!\!\!\!\!\!\!\!\!\,f_{\mathrm{dtb}}(\, r, \, e,\,p\,)
\end{align*}
\label{lma:totalCausalInfluences}
\end{lemma}
\textit{\textbf{Proof:} Firstly, for any relation $r$, the expression $\mathrm{C}_{r}$ determines (for the general case), the direct causal influence exchanged by the relation. When subtracted from the total disturbance $\mathrm{D}_r$ of a given experiment, $\mathrm{D}_r - \mathrm{C}_{r}$ quantifies the total noise. This is known because all disturbances that are not the result of causal influences are inevidably the result of noise. In some cases, this may be a negative value (i.e., when there is no disturbance on a relation that anticipates disturbance comprised of causal influence), and so the value is integrated into the expression $\mathrm{max}\,\{\,0, \, \mathrm{D}_r - \mathrm{C}_r \,\}$. Finally, the value is again subtracted from $\mathrm{D}_r$ to determine the direct causal influences in the experiment itself. It follows that the summation for all relations quantifies the total direct causal influences for the entire experiment.}$\hfill\blacksquare$

\subsection{Determining Contextuality By Causal Influences}

At this stage, it is possible to redefine the equation used to determine quantum-like contextuality. Prior to this work, \citet{pironio2003violations} states that for any Bell parameter $\mathcal{B}$ that violates the Bell inequalities by means of causal influences only, that the amount by which the violation is observed is exactly equivalent to the exchanged causal influences.

\begin{corollary}
For an experiment that determines contextuality, let the Bell parameter $\mathcal{B}$ generate its statistical correlation by means of causal influences \emph{only} (i.e., in the full absence of noise among experimental results). Also, let the weightings of said causal influences be totalled to the value $\mathrm{C}_{\mathrm{all}}$. Then it is known by \citet{pironio2003violations} that the amount $\mathcal{B}$ violates the statistical bound $\mathcal{B}_0$ is equivalent to $\mathrm{C}_{\mathrm{all}}$.
\begin{align*}
    \left| \mathcal{B} - \mathcal{B}_0 \right| = \mathrm{C}_{\mathrm{all}}
\end{align*}
\label{cly:BellParameterEqualsCausalInfluences}
\end{corollary}

Finally, by rearrangement of Corollary \ref{cly:BellParameterEqualsCausalInfluences} and integration of Lemma \ref{lma:totalCausalInfluences}, we obtain an expression that can determine contextuality in the presence of causal influences.
\begin{theorem} 
For any given experiment, let the Bell parameter $\mathcal{B}$ generate its statistical correlation against a statistical bound $\mathcal{B}_0$. Then, if the total disturbances for the experiment that constitute causal influences subtracted from the Bell parameter are greater than $\mathcal{B}_0$, the experiment determines contextuality.
\begin{align*}
    \mathcal{B} - \left(\;\sum_{r \,\in\, R}\,\mathrm{D}_r - \mathrm{max}\,\{\,0, \, \mathrm{D}_r - \mathrm{C}_r \,\}\right) \;>\; \mathcal{B}_{0}
\end{align*}
\label{thm:contextualityAmidstCausalInfluences}
\end{theorem}

By Theorem \ref{thm:contextualityAmidstCausalInfluences}, we can determine contextuality in the presence of causal influences. More importantly, this result allows us to correctly determine quantum-like contextuality within cognitive experiments.

\section{An Example Scenario}

In this section, an example is provided to convey how the extension of the combinatorial approach determines contextuality in the presence of causal influences. The example again considers the EPR framework, however with a separate probabilistic model that has the pair-wise joint distributions in Table \ref{tab:pairwiseJointIndivExpTrialsCombined}.

In the literature, the example is widely known as a `Popescu-Rohrlich' (P-R) box \citep{popescu1998causality}, and is known to be maximally contextual due to the degree that it's Bell parameter $\mathcal{B}$ violates the bound $\mathcal{B}_0$ of noncontextual hidden variable theories.

\begin{equationn}
The Bell parameter of Table \ref{tab:pairwiseJointIndivExpTrialsCombined} violates the bound of noncontextual hidden variable theories.
\begin{align*}
    \mathcal{B}_0 &\not\geq \mathcal{B}\\
    2 \;&\not\geq\; \underset{
 \begin{matrix}
 a \,\in\,\{ \,+1, \,-1 \,\}\\
 b \,\in\,\{ \,+1, \,-1 \,\}
 \end{matrix}}{\mathrm{max}} 
 \raisebox{-0.8em}{\scalemath{2}{|}}\;
 \mathrm{corr}_{\scalemath{0.75}{\begin{matrix}{}^{\mathrm{ipt}}\!A\,=\,+1\\[-0.25em]{}^{\mathrm{ipt}}\!B\,=\,+1\end{matrix}}} \;+\; 
 \mathrm{corr}_{\scalemath{0.75}{\begin{matrix}{}^{\mathrm{ipt}}\!A\,=\,+1\\[-0.25em]{}^{\mathrm{ipt}}\!B\,=\,-1\end{matrix}}}
 \\[-1.5em]
&\;\;\;\;\;\;\;\;\;\;\;\;\;\;\;\;\;\;\;\;\;\;\;\;\;\;\;\;\;\;\;\;\;\;+\;
\mathrm{corr}_{\scalemath{0.75}{\begin{matrix}{}^{\mathrm{ipt}}\!A\,=\,-1\\[-0.25em]{}^{\mathrm{ipt}}\!B\,=\,+1\end{matrix}}} \;+\; \mathrm{corr}_{\scalemath{0.75}{\begin{matrix}{}^{\mathrm{ipt}}\!A\,=\,-1\\[-0.25em]{}^{\mathrm{ipt}}\!B\,=\,-1\end{matrix}}}
\;-\; 2\mathrm{corr}_{\scalemath{0.75}{\begin{matrix}{}^{\mathrm{ipt}}\!A\,=\,a\\[-0.25em]{}^{\mathrm{ipt}}\!B\,=\,b\end{matrix}}}\; \raisebox{-0.8em}{\scalemath{2}{|}}\\
2 \;&\not\geq\; \left| 1 - 1 + 1 + 1 - 2(-1) \right| \;\;\; \mathrm{given} \;\;\; a = +1 \;\; b = -1\\
2 \;&\not\geq\; 4
\end{align*}
\label{eq:supra}
\end{equationn}

Earlier in this article, it was mentioned that results such as the above may be due to causal influences, and may not be truly contextual in nature. By the relevant techniques, an experimental protocol is now detailed.

\begin{figure}[H]
\begin{center}
     \includegraphics[width=0.55\textwidth]{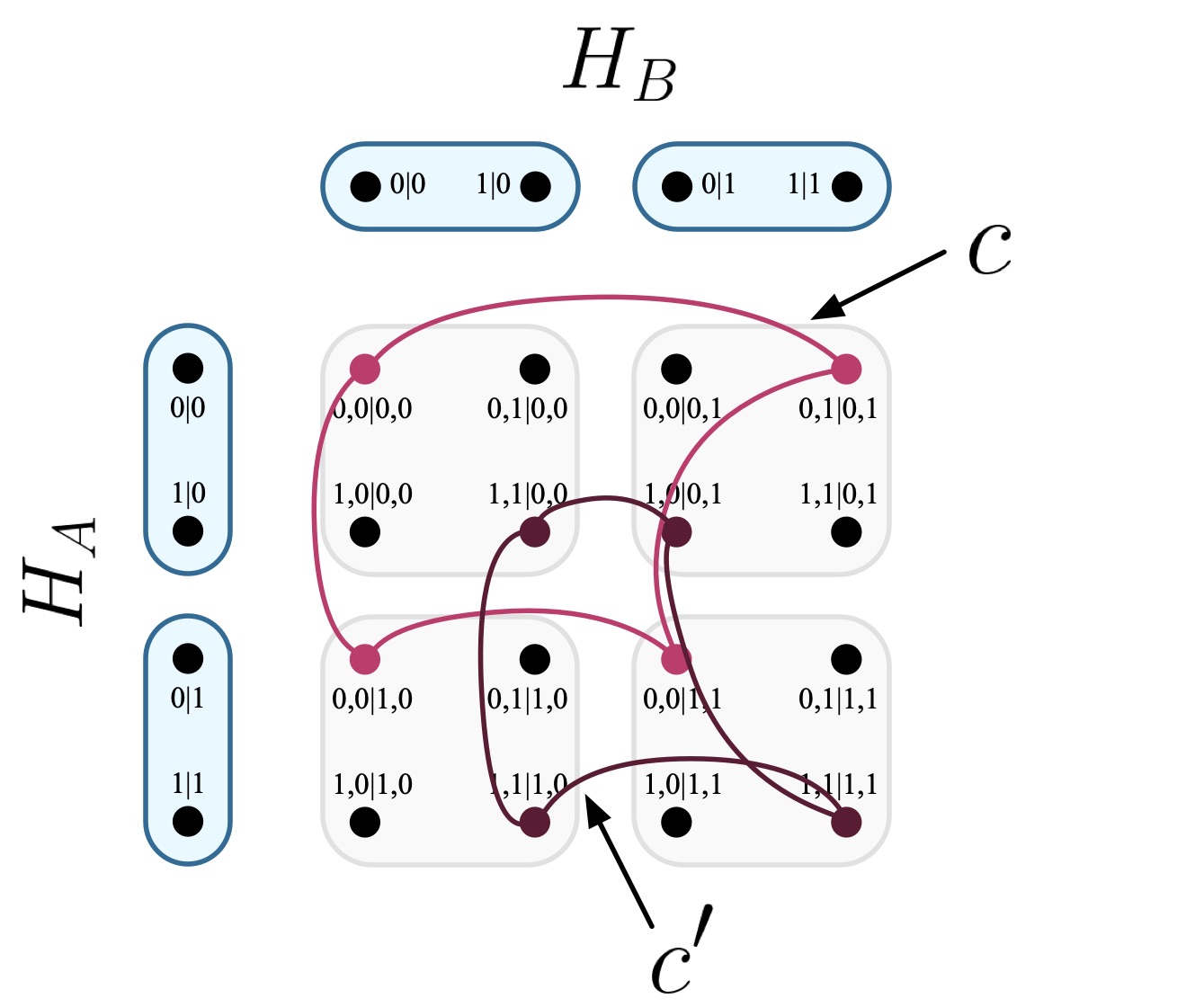}
\end{center}
	\caption{The Cliques $c$ And $c'$ Visualised On Cartesian Product Of Contextuality Scenarios $H_A$ And $H_B$}
\label{fig:CliquesOnContextualityScenario}
\end{figure}

Let it be such that the experimental trials that inform the probabilistic model of Table \ref{tab:pairwiseJointIndivExpTrialsCombined} have two equally likely deterministic models, which correspond to the cliques seen in Figure \ref{fig:CliquesOnContextualityScenario}.

Specifically, the cliques would each be attributed equal weightings by the set $q$.

\begin{definition}
The set $q$, conveyed by the two comprising weights that correspond to the cliques in Figure \ref{fig:CliquesOnContextualityScenario}: 
\begin{align*}
    q_{c} = 0.5, \;\;\;\;\;\; q_{c'} = 0.5
\end{align*}
\end{definition}

From either clique, it can be speculated that the outcomes are not invariant with respect to the measurements, and are thus dependent upon their selections. This is evaluated by means of mapping the measurements to a causal model, as previously visualised in Figure \ref{fig:CausalModelAlternative1}.

Next, all relations upon the causal model are evaluated; as mentioned in Section \ref{sec:causalInfluences}, these correspond to the causal influences that may be present in the model. Of interest, the following relations are considered.

\begin{definition}
Four relations of set $R$ that are of significance to possible causal influences within the example's causal model:
\begin{align*}
    r\;&=\; \{\,(\,f_{\mathrm{vtc}}(\,do(\, X_1 = +1, \; \mathrm{X}_2\,=\,+1 \,)\,), \,f_{\mathrm{vtc}}(\,\mathrm{X}_3 \,= \,-1,\,\mathrm{X}_4 \,= \,+1\,)\,)\,\}\\
r'\;&=\; \{\,(\,f_{\mathrm{vtc}}(\,do(\, X_1 = +1, \; \mathrm{X}_2\,=\,-1 \,)\,), \,f_{\mathrm{vtc}}(\,\mathrm{X}_3 \,= \,-1,\,\mathrm{X}_4 \,= \,+1\,)\,)\,\}\\
r''\;&=\; \{\,(\,f_{\mathrm{vtc}}(\,do(\, X_1 = -1, \; \mathrm{X}_2\,=\,+1 \,)\,), \,f_{\mathrm{vtc}}(\,\mathrm{X}_3 \,= \,-1,\,\mathrm{X}_4 \,= \,+1\,)\,)\,\}\\
r'''\;&=\; \{\,(\,f_{\mathrm{vtc}}(\,do(\, X_1 = -1, \; \mathrm{X}_2\,=\,-1 \,)\,), \,f_{\mathrm{vtc}}(\,\mathrm{X}_3 \,= \,-1,\,\mathrm{X}_4 \,= \,+1\,)\,)\,\}
\end{align*}
\label{def:4Rs}
\end{definition}

For the first relation $r$, we wish to determine the causal influence of observing $\mathrm{X}_3 = -1$, given that $\mathrm{X}_1$ and $\mathrm{X}_2$ are both fixed to the outcomes $+1$. To do this, we firstly calculate the edges that correspond to relation $r$ by Definition \ref{def:ER}.

\begin{definition}
For relation $r$ from Definition \ref{def:4Rs}, the calculation of edges from the necessary measurement protocol that correspond to the set $E_r$:
\begin{align*}
    E_r \;&:=\; \left \{ \,e \,:\, e\, \in \,E_{\,H_i \,\rightarrow \,H_j\,} \;\; \mathrm{where} \;\; x \subseteq H_j,\; y \subseteq H_i, \;\; r \,:=\, \{\,(\,x, \,y\,)\,\}, \;\; \mathrm{and}\;\; \left \{\, e \, \cap \, \times^{|r|}_{i=1} r \,\right \} \neq \{\,\varnothing\,\} \, \right \}\\
\;&:=\; \{ \,e \,:\, e\, \in \,E_{\,H_A \,\rightarrow \,H_B\,} \;\; \mathrm{where} \;\; \{( \,v_{0|0}\, )\} \subseteq H_A,\; \{( \,v_{0|1}\, )\} \subseteq H_B, \;\;\\
& \;\;\;\;\;\;\;\;\;\;\;\;\;\;\;\;\;\;\;\;\;\;\;\;\;\;\;\;\;\;\;\;\;\;\; r \,:=\, \{\,(\,v_{0|0}, \,v_{0|1}\,)\,\}, \;\; \mathrm{and}\;\; \left \{\, e \, \cap \, v_{0,0|0,1} \,\right \} \neq \{\,\varnothing\,\} \, \}\\
\;&:=\; \left \{(\, v_{0,0|0,1}, \, v_{1,0|0,1}, \, v_{0,1|1,1}, \, v_{1,1|1,1} \,)\right \}
\end{align*}
Note: In calculating set $E_r$, the comprising values of $x$ and $y$ have been derived by the following additional equations:
\begin{align*}
x \;&:=\; f_{\mathrm{vtc}}(\,do(\, X_1 = +1, \; \mathrm{X}_2\,=\,+1 \,)\,)\\
x \;&:=\; \{( \,v_{0|0}\, )\}\\[1ex]
y \;&:=\; f_{\mathrm{vtc}}(\,\mathrm{X}_4 \,= \,+1, \,\mathrm{X}_3 = -1\,)\\
y \;&:=\; \{(\,v_{0|1}\,)\}
\end{align*}
\label{def:getER}
\end{definition}

\begin{figure}[H]
\begin{center}
     \includegraphics[width=0.3\textwidth]{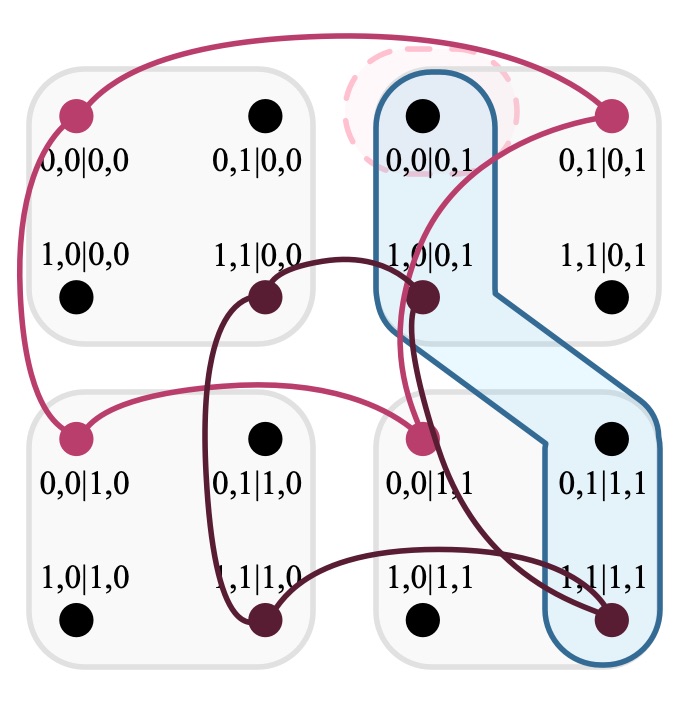}
\end{center}
	\caption{Intersection Of Cliques, A Hyperedge Of The F-R Product, And A Relation}
	\textit{Note: The hyperedge returned by Definition \ref{def:getER} is highlighted in blue; the relation $r$ is highlighted in light red.}
\label{fig:DiagramIsolatedEdge}
\end{figure}

Having determined $E_r$ in Definition \ref{def:getER}, only a single edge of the measurement protocol $E_{H_A \rightarrow H_B}$ within the F-R product has been returned. This is visualised in Figure \ref{fig:DiagramIsolatedEdge}.

Now it is possible to calculate the causal influence variable $C_r$ by Definition \ref{def:CausalVariable}. As specified, there is a requirement that two separate experimental tests are conducted prior to its calculation, to derive the necessary probabilistic models that correspond to fixing the causal variables to the outcomes of relation $r$ (i.e., $\mathrm{Pr}(\,\mathrm{X}_4 = +1\; | \; do(\,\mathrm{X}_1 = +1, \,\mathrm{X}_2 = +1\,), \,\mathrm{X}_3 = -1\,)$), as well as fixing the causal variables \textit{not} to the outcomes of relation $r$ (i.e., $\mathrm{Pr}(\,\mathrm{X}_4 = +1\; | \; do(\,\mathrm{X}_1 = +1, \,\mathrm{X}_2 = -1\,), \,\mathrm{X}_3 = -1\,)$). Coincidentally, the latter case is exactly $r'$, as detailed in Definition \ref{def:4Rs}. For both relations, the causal variables are significantly fixed, and have separate probabilistic models ($p$ for $r$, and $p'$ for $r'$) informed by their respective experiments. The results, as sampled under the causal interventions are given in Table \ref{tab:ProbDistCorr}.

\setcounter{subfigure}{0}
\begin{table}[H]%
    \setstretch{1.75}
    \centering
    \subfloat[\centering $\mathrm{X}_4$, given $ do(\,\mathrm{X}_1 = +1, \,\mathrm{X}_2 = +1\,)$  and $\mathrm{X}_3 = -1$]{{\begin{tabular}{ c }
        $\begin{matrix}[ c c ] \mathrm{X}_4 = +1 & \mathrm{X}_4 = -1\end{matrix}$ \nonumber \\
            $\begin{pmatrix}[cc] \;\; 0.00 \;&\; 1.00\;\; \end{pmatrix}$
    \end{tabular}}}%
    \qquad
    \subfloat[\centering $\mathrm{X}_4$, given $ do(\,\mathrm{X}_1 = +1, \,\mathrm{X}_2 = -1\,)$  and $\mathrm{X}_3 = -1$]{{\begin{tabular}{ c }
        $\begin{matrix}[ c c ] \mathrm{X}_4 = +1 & \mathrm{X}_4 = -1\end{matrix}$ \nonumber \\
            $\begin{pmatrix}[cc] \;\; 1.00 \;&\;0.00\;\; \end{pmatrix}$
    \end{tabular}}}%
    \caption{Probabilistic Distributions Corresponding To Observation Of $\mathrm{X}_4$ Under Causal Intervention}%
    \label{tab:ProbDistCorr}%
\end{table}

With knowledge of $p$ and $p'$, $C_r$ is calculated by determining the difference in disturbance between the relations $r$ and $r'$, given their respective probabilistic models. It is known that the F-R product calculates said disturbance, and that by relation $r$, the edge given in Figure \ref{fig:DiagramIsolatedEdge} is the only edge that qualifies for the relation; this is the same for the relation $r'$. What differs between them is that clique $c'$ does not intersect relation $r$, which then influences the result of $C_r$.

\begin{definition}
The causal influence $C_r$ determined for the relation $r$, specifically for the causal influence that the configuration of $\mathrm{X}_1 = +1$ and $\mathrm{X}_2 = +1$ exert upon $\mathrm{X}_4$ when $\mathrm{X}_3 = -1$.
\begin{align*}
\mathrm{C}_r\;&= \!\!\!\!\!\!\!\!\!\!\!\!\!\!\!\!\!\!\!\!\!\!\sum_{\;\;\;\;\;\;\;\;\;\;\;\;\;\;\;e\, \in \,E(\,\bigotimes^{n}_{i=1}H_i\,)} \!\!\!\!\!\!\!\!\!\!\!\!\!\!\!\!\!\!\left|\,f_{\mathrm{dtb}}(\, r, \, e,\,p\,) - f_{\mathrm{dtb}}(\, r', \, e,\,p'\,)\,\right|\\
&=\; \left|\,0 \;-\;q_{c'}\!\left|\left|\left\{ \, c' \, \cap \, e\, \right\}\right| - 1\right| \right|\\
&=\; \left|\,0 \; - \; 0.5\right|\\
&=\; 0.5
\end{align*}
\end{definition}
The same calculation is obtained for all other relations, which for the example reveals the following values:
\begin{definition}
The causal influence variables, as calculated for all relations of the example:
\begin{align*}
C_r = 0.5, \;\;\;\;\;\; C_{r'} = 0.5, \;\;\;\;\;\; C_{r''} = 0.5, \;\;\;\;\;\; C_{r'''} = 0.5
\end{align*}
\end{definition}

By Equation \ref{eq:supra}, the Bell parameter $\mathcal{B}$ is known to equal $4$; also by Equation \ref{def:bellCHSHequaltiies}, the statistical bound on noncontextual hidden variables $\mathcal{B}_{0}$ is known to equal $2$. In Theorem \ref{thm:contextualityAmidstCausalInfluences}, the value $D_r$ is known to calculate the causal influence of each relation $r$ under the pretense of no causal intervention. For the example, this so happens to be equivalent to the value of causal variable (i.e. $D_r = C_r$). It follows that the cancellation of values is reflected in the calculation of the theorem below:

\begin{equationn}
The calculation of Theorem \ref{thm:contextualityAmidstCausalInfluences} for the example scenario given in this section.
\begin{align*}
\mathcal{B} - (\;\sum_{r \,\in\, R}\,\mathrm{D}_r - \mathrm{max}\,\{\,0, \, \mathrm{D}_r - \mathrm{C}_r \,\}) \;&\not>\; \mathcal{B}_{0}\\
4 - ( (\mathrm{D}_r - \mathrm{max}\,\{\,0, \, \mathrm{D}_r - \mathrm{C}_r \,\}) + (\mathrm{D}_{r'} - \mathrm{max}\,\{\,0, \, \mathrm{D}_{r'} - \mathrm{C}_{r'} \,\})\;\;\;\;\;\;\;\;\;\;\;\;& \\
+ (\mathrm{D}_{r''} - \mathrm{max}\,\{\,0, \, \mathrm{D}_{r''} - \mathrm{C}_{r''} \,\}) + (\mathrm{D}_{r'''} - \mathrm{max}\,\{\,0, \, \mathrm{D}_{r'''} - \mathrm{C}_{r'''} \,\}) ) \;&\not>\; 2\\
4 - ((0.5 - 0) + (0.5 - 0) + (0.5 - 0) + (0.5 - 0)) \;&\not>\; 2\\
2 \;&\not>\; 2
\end{align*}
\label{eq:thmNotCNTX}
\end{equationn}

As can be seen from the result of Equation \ref{eq:thmNotCNTX}, the causal influences exchanged between the measurements and outcomes of the respective parties cancel out any indication that the Bell parameter is quantum-like contextual. This concludes the example scenario of how the techniques developed in this article may be applied to more adequately determine quantum-like contextuality.

\section{Conclusion}

This article has developed and integrated a set of modelling techniques to address the challenges of determining quantum-like contextuality in the presence of causal influences. It has achieved this by firstly addressing the challenge of providing meaningful results to experimentation when disturbances are present, by combination of the F-R product of \citet{acin2015combinatorial}'s combinatorial approach and \citet{chaves2015unifying}'s causal influence formula. The results of this article have also addressed the second challenge of providing a sensitive treatment to the convex decomposition of the probabilistic model associated with an experiment by means of the combinatorial approach's WFPN. This further ensures that hidden causal influences are accounted for within experimental results, as previously recommended by \citet{atmanspacher2019contextuality}. In addressing these challenges, it has furthermore become possible to derive a novel theorem (see Theorem \ref{thm:contextualityAmidstCausalInfluences}) that reasonably adjusts the statistical bound of noncontextual hidden variable theories, allowing for a theoretically consistent determination of contextuality. 

Beyond the main objectives of the article, we have also clarified incorrect causal assumptions introduced by the usage of the ND condition within experimentation for determining contextuality. As detailed in Section \ref{sec:associationNDBad}, it was found that the ND condition can incorrectly classify noise within experimental results as causal influences, which inhibits a meaningful interpretation of experimental results. The article also details how measurements and outcomes (as they are formalised in the combinatorial approach) may be related to canonical causal models. This has been achieved by the necessary mapping functions of causal variables and effects to contextuality scenarios and relations respectively, as detailed in Section \ref{sec:ModellingCombinatorial}. Lastly, the article has provided a comprehensive example of how the theoretical contributions may be applied. A protocol detailing the experimental steps has been developed, such that a cognitive modeller may apply them to arrive at a meaningful result. While the example is communicated in the manner of the EPR framework, the theory is generally distilled to be applicable within systems of arbitrarily many outcomes, measurements or parties.

In terms of limitations, it is perceived that this work has only considered integrating the absolutely necessary causal modelling techniques required to produce viable experimental results. We recognise that contemporary approaches to modelling contextuality do not include rigorous causal analyses by nature, and this was evident by the absence of such techniques within the combinatorial approach. However it seems necessary to interrogate causal approaches to study contextuality, which we perceive will further our understanding of the phenomenon. Aside from this, we believe it would also be necessary to adapt Theorem \ref{thm:contextualityAmidstCausalInfluences} to a linear program that can more generally calculate the statistical bound for any experimental protocol, given the presence of causal influences. It is known in the literature that the statistical bound of the Bell inequalities form a convex polytope of all probabilistic models that adhere to noncontextual hidden variable theories. In saying this, while the methods described in this paper have served the derivation of the statistical bounds necessary to achieve Theorem \ref{thm:contextualityAmidstCausalInfluences}, it is nevertheless possible to extend the findings even further to a more straightforward solution.

\section*{Conflict of Interest Statement}

The authors declare that the research was conducted in the absence of any commercial or financial relationships that could be construed as a potential conflict of interest.

\section*{Author Contributions}

In terms of contribution, Dr Obeid and Prof Bruza were responsible for the development of the theories necessary to achieve the fundamental results of the paper. Dr Obeid was primarily responsible for drafting the visual materials and example scenario. All authors were responsible for the proof-reading, grammatical structure, and communication of findings.

\section*{Funding}

This research has been partially funded by the Unitary Fund, and the Asian Office of Aerospace Research and Development (AOARD) grant: FA2386-17-1-4016.

\section*{Acknowledgments}

In developing the main findings of this article, we'd like to thank Dr Elie Wolfe for the wealth of guidance he has provided.

\bibliographystyle{plainnat}
\bibliography{references.bib}

\end{document}